\documentclass[iop]{emulateapj}
\usepackage{multirow}
\usepackage{gensymb}
\usepackage{amsmath, amssymb}
\usepackage{booktabs}
\newcommand{\nii}{[\ion{N}{2}]}
\newcommand{\ha}{$\mathrm{H{\alpha}}$}
\newcommand{\hb}{$\mathrm{H{\beta}}$}
\newcommand{\oiii}{[\ion{O}{3}]}
\newcommand{\oii}{[\ion{O}{2}]}
\newcommand{\msun}{\mathrm{M}_{\odot}}
\newcommand{\nha}{[\ion{N}{2}]/$\mathrm{H\alpha}$}
\newcommand{\ohb}{[\ion{O}{3}]/$\mathrm{H\beta}$}

\begin{document}

\title{Metal deficiency in cluster star-forming galaxies at $\lowercase{z}=2$}
\author{\sc Francesco Valentino \altaffilmark{1,2}}
\altaffiltext{1}{Laboratoire AIM-Paris-Saclay, CEA/DSM-CNRS-Universit\'{e} Paris Diderot, Irfu/Service d'Astrophysique, CEA Saclay, Orme des Merisiers, F-91191 Gif sur Yvette, France}
\altaffiltext{2}{francesco.valentino@cea.fr}
\author{\sc Emanuele Daddi\altaffilmark{1}}
\author{\sc Veronica Strazzullo\altaffilmark{1,3}}
\altaffiltext{3}{Department of Physics, Ludwig-Maximilians-Universit\"{a}t, Scheinerstr. 1, 81679 M\"{u}nchen, Germany}
\author{\sc Rapha\"{e}l Gobat\altaffilmark{1,4}}
\altaffiltext{4}{School of Physics, Korea Institute for Advanced Study, Heogiro 85, Seoul 130-722, Republic of Korea}
\author{\sc Masato Onodera\altaffilmark{5}}
\altaffiltext{5}{Institute for Astronomy, ETH Z\"{u}rich Wolfgang-Pauli-strasse 27, 8093 Z\"{u}rich, Switzerland}
\author{\sc Fr\'ed\'eric Bournaud\altaffilmark{1}}
\author{\sc St\'{e}phanie Juneau\altaffilmark{1}}
\author{\sc Alvio Renzini\altaffilmark{6}}
\altaffiltext{6}{INAF-Osservatorio Astronomico di Padova Vicolo dell'Osservatorio 5, I-35122 Padova, Italy}

\author{\sc Nobuo Arimoto\altaffilmark{7,8}}
\altaffiltext{7}{Subaru Telescope, National Astronomical Observatory of Japan 650 North A'ohoku Place, Hilo, HI 96720, USA}
\altaffiltext{8}{Graduate University for Advanced Studies, 2-21-1 Osawa, Mitaka, Tokyo, Japan}

\author{\sc Marcella Carollo\altaffilmark{5}}

\author{\sc Anita Zanella\altaffilmark{1}}

\begin{abstract}
We investigate the environmental effect on the metal enrichment of star-forming galaxies (SFGs) in the farthest spectroscopically confirmed and X-ray detected cluster, CL~J1449+0856 at $z=1.99$. We combined \textit{HST}/WFC3~G141 slitless spectroscopic data, our 13-band photometry, and a recent Subaru/MOIRCS near infrared spectroscopic follow-up to constrain the physical properties of SFGs in CL~J1449+0856 and in a mass-matched field sample. After a conservative active galactic nuclei (AGN) removal, stacking individual MOIRCS spectra of $6$ ($31$) sources in the cluster (field) in the mass range $10 \leq \mathrm{log(M/\msun}) \leq 11$, we find a $\sim4\sigma$ significant lower \nha\ ratio in the cluster than in the field. Stacking a subsample of $16$ field galaxies with \hb\ and \oiii\ in the observed range, we measure a \ohb\ ratio fully compatible with the cluster value. Converting these ratios into metallicities, we find that the cluster SFGs are up to $0.25\,\mathrm{dex}$ poorer in metals than their field counterparts, depending on adopted calibrations. The low metallicity in cluster sources is confirmed using alternative indicators. Furthermore, we observe a significantly higher \ha\ luminosity and equivalent width in the average cluster spectrum than in the field. This is likely due to enhanced specific star formation rate, even if lower dust reddening and/or an uncertain environmental dependence of the continuum-to-nebular emission differential reddening may play a role. Our findings might be explained by the accretion of pristine gas around galaxies at $z=2$ and from cluster-scale reservoirs, possibly connected with a phase of rapid halo mass assembly at $z>2$ and of high galaxy merging rate.
\end{abstract}	


\section{Introduction}
The evolution of galaxies is regulated by the complex interplay of multiple physical mechanisms. Distinguishing the influence of external environmental effects from internal factors is crucial to reach a comprehensive understanding of these systems. In this perspective, galaxy clusters offer the perfect occasion to disentagle this situation, comparing samples of field and extreme overdensity galaxies at fixed mass. In the local Universe the influence of the strongest overdensities is manifest in well-known relations, such as the systematic variation of morphological type, luminosity, surface brightness, star formation rate (SFR), and colors with density \citep[e.g.,][]{dressler_1980,gomez_2003,baldry_2004,balogh_2004,hogg_2004,blanton_2005}. As a result, local virialized and massive clusters are centrally dominated by massive, red, and passive early-type galaxies, while blue star-forming galaxies (SFGs) are preferentially located in the cluster outskirts and in the field. A key to decipher the origin of the observed local environmental trends is the study of high redshift cluster galaxies as compared to the field. Unlike the extended and increasing statistics of well studied clusters at $z<1.5$ in the literature, only a handful of clusters above this redshift are confirmed today \citep[e.g.,][]{andreon_2009,papovich_2010,gobat_2011,fassbender_2011,santos_2011,stanford_2012,zeimann_2012,muzzin_2013}. Studying the redshift interval above $z>1.5$ is crucial, as we approach the era when the first massive clusters begin to emerge and an epoch where galaxies were still assemblying a large fraction of their stellar mass through active star formation \citep{daddi_2007}. In addition to the low statistics, the mentioned properties which designate an evolved cluster at $z=0$ become progressively blurred at increasing redshift, making it difficult to fully characterize the evolutionary stage of overdensities and, consequently, to quantify their effect on galaxy evolution. Despite these difficulties, sustained efforts have been and continue to be made to detect, confirm and characterize high redshift clusters. In particular, the recent dramatic improvement in near-IR multi-object spectrographs has opened the way to study the physical properties of ionized gas in SFGs through a set of emission lines well studied in local objects, such as \hb, \oiii$\lambda5007\,\mathrm{\AA}$ (hereafter \oiii), \ha, and \nii$\lambda6584\,\mathrm{\AA}$ (hereafter \nii). These lines give also access to the gas-phase metallicity in SFGs, if properly calibrated. A Mass-Metallicity Relation (MZR) has been shown to be in place from $z=0$ \citep{tremonti_2004} up to redshift $\sim3-4$ \citep[][and others]{savaglio_2005, erb_2006, troncoso_2014, zahid_2013, wuyts_2014}, indicating that more massive galaxies are also more metal rich at almost any epoch. As recent modelling suggests \citep[e.g.,][]{dave_2012,lilly_2013}, this relation may result from secular metal enrichment of the gas through stellar winds from young stars, modulated by galactic outflows and inflows and by the formation of subsequent generation of stars. At increasing redshift, the gas-phase metallicity in SFGs is observed to decrease, but our knowledge of possible effects of the surrounding environment on metal enrichment is still uncertain. In the local Universe the environmental effect seems to be limited, if at all present \citep{mouhcine_2007, cooper_2008, ellison_2009, hughes_2013}, and recent studies at high redshifts have focused only on few protoclusters at $z>2$ \citep{kulas_2013, shimakawa_2014}. %., with the detection of a metal enhancement in SFGs residing in these overdensities. \\
In this work we present results relative to the farthest spectroscopically confirmed X-ray detected cluster discovered to date, CL J1449+0856 at $z=1.99$ \citep[hereafter G11, G13]{gobat_2011, gobat_2013}. The presence of a dominant population of red, massive, and passive galaxies in its core \citep[hereafter S13]{strazzullo_2013}, coupled with the X-ray detection, places CL J1449+0856 in a relatively evolved phase compared with other known structures at the same epoch, making it the potential progenitor of a massive local cluster (G11). These features physically distinguish this overdensity from lower halo mass, SFG dominated, rapidly assembling protoclusters at similar or higher redshifts \citep[e.g.,][]{steidel_2005,kodama_2013} and potentially these intrinsically different structures may give rise to different effects on their host galaxies. We present here a recent Subaru/Multi-Object InfraRed Camera and Spectrograph (MOIRCS) follow-up of the star-forming population in CL J1449+0856, for which we primarily measured \ha\ and \nii\ emission lines. Incorporating previous information about \oiii\ and \hb\ from \textit{HST}/WFC3 G141 slitless spectroscopy \citep{gobat_2013}, we can estimate the metallicity through the $N2=\mathrm{log}($\nii$/$\ha$)$ and $O3N2=\mathrm{log}[($\oiii$/$\hb$)/($\nii$/$\ha$)]$ indicators \citep{alloin_1979}, exploring different calibrations proposed in recent literature \citep{pettini_2004, steidel_2014}. We compare the MZR and other interstellar medium (ISM) properties of SFGs in CL J1449+0856 with a mass-matched field sample at comparable redshift, allowing for a direct probe of the environmental effects of relatively evolved overdensities on SFGs at $z=2$.\\
This paper is organized as follows. In Section \ref{sec:data_sample} we describe the sample selection and the near-IR spectroscopic and ancillary data used for the analysis. In Section \ref{sec:methodologies} we present the full photometric and spectroscopic analysis of the dataset along with the derived ISM physical condition through currently used line diagnostic diagrams. We show the results about the MZR in Section \ref{sec:mzr} and we discuss potential implications in Section \ref{sec:discussion}. Section \ref{sec:conclusion} summarizes the main findings of this work. Additional technical remarks are reported in the Appendix. Throughout all the paper we adopt a flat $\Lambda$CDM cosmology with $\Omega_{\mathrm{m}} = 0.3$, $\Omega_{\mathrm{\Lambda}} = 0.7$ and $H_0 = 70\,\mathrm{km\,s^{-1}\,Mpc^{-1}}$, and a Salpeter initial mass function \citep[IMF,][]{salpeter_1955}. When necessary we converted results from literature obtained with other IMFs to a Salpeter IMF.

\section{Data and sample selection}
\label{sec:data_sample}
\begin{figure}
  %\hspace*{-2cm}
  \includegraphics[width=0.5\textwidth, angle=0]{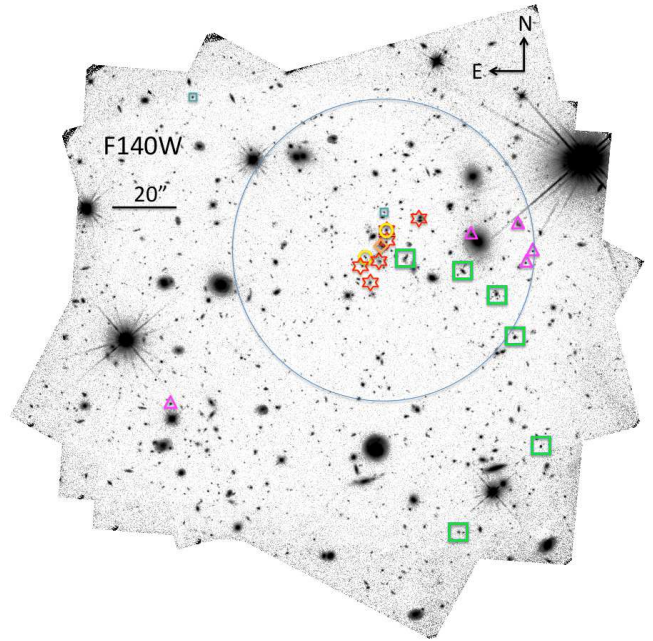}
  \caption{Deep F140W image of CL J1449+0856 and its spectroscopically confirmed members in the field observed with WFC3 (G13). Red stars mark passive galaxies. Yellow circles indicate X-ray detected AGNs. Green squares indicate SFGs in the mass range $10 \leq \mathrm{log(M/\msun}) \leq 11$ and blue squares indicate SFGs in the mass range $ \mathrm{log(M/\msun}) < 10$, both targeted and detected with MOIRCS. Purple triangles indicate other SFGs not targeted with MOIRCS. The orange diamond shows the assemblying brightest central galaxy. The blue solid circle represents the putative $\mathrm{R_{200}}\sim0.4\,\mathrm{Mpc}$ radius (physical, G13).}
  \label{fig:cluster_map}
\end{figure}
Galaxy populations in CL J1449+0856 were investigated in G11, G13, and S13. The cluster is spectroscopically confirmed with currently 27 members identified with VLT/VIMOS and FORS2, and WFC3 spectroscopy (G11, G13). Most spectroscopic redshifts in the field of CL J1449+0856 come from the WFC3 G141 spectroscopic follow-up (Figure \ref{fig:cluster_map}), with 140 redshift determinations over a $\sim 4\,\mathrm{arcmin}^2$ area, based on emission lines (typically \oii$\lambda3727\,\mathrm{\AA}$, [\ion{Ne}{3}], \hb,  \oiii\ at $z\sim2$) or on continuum breaks in the spectral range $1.1 - 
1.7\, \mu\mathrm{m}$ (full details can be found in G13). These include $68$ \oiii\ emitters, $17$ of which belong to the cluster.
CL J1449+0856 was also imaged at wavelengths from X-ray to radio (G13). In this work, we used the same photometric catalogs as in 
S13, including optical/NIR photometry in $13$ passbands from \textit{U} to $4.5\,\mu\mathrm{m}$. Sources were detected in the WFC3/F140W band, and photometry was measured with SExtractor \citep{bertin_1996}, as well as with GALFIT \citep{peng_2002,peng_2010} modelling. Based on photometric redshifts determined on such photometry, a sample of candidate cluster members was identified in the cluster central region, virtually complete at $\mathrm{M}\gtrsim10^{10}\,\mathrm{M_{\odot}}$, although affected by significant contamination especially below $10^{10}\mathrm{M_{\odot}}$. Galaxies were also broadly classified as ``passive'' or ``star-forming'' based on restframe UVJ colors \citep{wuyts_2007,williams_2009} and spectral energy distribution fitting (SED, S13). In this work, we focus on SFGs in the mass range $10 \leq \mathrm{log(M/M_{\odot})} \leq 11$. The full sample of cluster galaxies in the F140W-based catalog includes $6$ spectroscopically confirmed and $2$ candidate star-forming cluster members in this mass range. \\
For our MOIRCS near-IR follow-up, we selected a sample of $110$ objects. These included $76$ sources in CL 1449+0856 field, where we gave the highest priority to WFC3 spectroscopically confirmed star-forming members ($10$ objects: $2$ with $\mathrm{M} < 10^{10}\,\mathrm{M_{\odot}}$, $6$ with $10^{10}\,\mathrm{M_{\odot}} \leq \mathrm{M} \leq 10^{11}\,\mathrm{M_{\odot}}$, and $2$ with $\mathrm{M}>10^{11}\,\mathrm{M_{\odot}}$) and to star-forming objects from the pool of candidate members according to their probability of belonging to the cluster (S13) and irrespectively of their mass. We note here that cluster SFGs were not specifically selected to be \oiii\ emitters. The $2$ candidates in the mass range $10 \leq \mathrm{log(M/M_{\odot})} \leq 11$ were not observed due to geometrical constraints in slit positioning. In the area covered by WFC3 we selected $13$ \oiii\ emitters not belonging to the overdensity, which became part of the field control sample at $z\sim2$. Outside the WFC3 field, where no spectroscopy was available, we selected field objects with $z_{\mathrm{phot}} \geq 2$ with the highest chances to detect \ha, i.e. with an estimated \ha\ flux $\geq 3 \times 10^{-17}\,\mathrm{erg\, cm^{-2}\,s^{-1}}$ from the SED--based SFR and reddening estimates (see Section \ref{sec:methodologies}). Finally, to further extend our field control sample, we observed $34$ \textit{BzK}-SFGs \citep{daddi_2004} with $z_{\mathrm{phot}}\sim2$ with an estimated \ha\ flux $\geq 3 \times 10^{-17}\,\mathrm{erg\, cm^{-2}\,s^{-1}}$ in the COSMOS field \citep{scoville_2007}. A posteriori the predicted \ha\ was $\sim25$\% lower than the measured flux for these field \ha--selected sources, probably due to Malmquist bias. We note here that, even if the total integration time over the COSMOS field is shorter than over the cluster field, this does not substantially impact the main results of this work, based on the stacking of sources (see Section \ref{sec:stacking}). The objects in the COSMOS field contributed to $\sim30$\% of the total number of field sources in the final stacked spectrum ($10/31$) and reached \ha\ fluxes comparable with the dimmest sources in CL 1449+0856 field ($3.2 \times 10^{-17}$ and $3.3 \times 10^{-17} \,\mathrm{erg\, cm^{-2}\,s^{-1}}$ at $>5\sigma$ in the cluster and COSMOS field, respectively).

\subsection{Subaru/MOIRCS spectroscopy}
We carried out near-IR spectroscopy with MOIRCS at the Subaru Telescope \citep{ichikawa_2006}. Two Hawaii-2 $2048\times2048$ detectors cover the $4'\times 7'$ FoV and up to 40 slits can be placed within the inner $6'$ diameter circular region. We used the $HK500$ grism with $0.7"$ wide slits, which provides a resolving power $R\simeq 500$ along the $13000-23000\,\mathrm{\AA}$ spectral range. %The slit length has been changed from slit to slit accordingly to the spatial constraints on the slit displacement, but always assuring the conditions to effectively subtract the sky. 
A total of three masks were designed, two for the CL J1449+0856 field and one for the COSMOS field. %centered on the overdensity candidate at $z\sim2.2$ (Strazzullo et al., in prep.)
The observations were carried out in a single run of three consecutive nights in April, 2013. A sequence of $600\,\mathrm{s}$ integrations was taken with a standard ABAB $1.5"$ dithering pattern. Calibration frames of an $\mathrm{A0V}$ standard star and dome flat fields were taken at the beginning or end of each night. We integrated the images for a total of $7.3, 6.7, 3.3\,\mathrm{hr}$ on 38, 38, and 34 galaxies for Mask 1 and 2 on CL J1449+0856 and Mask 3 in the  COSMOS field, respectively, with a mean seeing of $\sim 0.6"$ during three clear nights. The observation plan is summarized in Table \ref{tab:observation_log}.\\
We reduced the data with the MCSMDP pipeline\footnote{http://www.naoj.org/Observing/DataReduction/} \citep{yoshikawa_2010} combined with custom IDL scripts. First, the data were flat-fielded employing dome-flat frames collected in the same configuration of science frames, and bad pixels and other detector defects were removed using masks provided in the pipeline. Then, cosmic rays were removed combining each A frame with the corresponding dithered B image. %This subtraction is not always satisfactory, thus an extra correction is applied introducing in the final combination of AB pairs a mild $\sigma$-clipping, which erases the remaining cosmic rays not affecting the brightest lines, as we have visually checked spectrum by spectrum. 
The sky subtraction was automatically performed subtracting each B frame to the corrisponding A image. Then the distortion introduced by the detectors was corrected using the coefficients used in the MOIRCS imaging reduction package. %The spectra on the detectors are slightly tilted at this step: thus, a rotation is applied to bring them parallel to the dispersion axis. 
Each 2D spectrum was then cut from every global frame and wavelength calibrated on a grid of bright OH-airglow lines \citep{rousselot_2000}, with an uncertainty of half of a pixel, i.e. $\sim3.5\,\mathrm{\AA}$. %Once aligned the sky lines to the spatial axis by means of the \texttt{transform} task in IRAF, we have added all the AB frames using the \texttt{imcombine} task, taking into account the proper offsets among the frames as measured from the position of a reference star positioned in each mask. 
We co-added all the 2D spectra, down-weighting the frames taken in worse atmospheric conditions to minimize the effect of variable seeing during the observing run. %Then we have extracted the 1D spectra using the \texttt{apall} task and flux-calibrated them dividing by the sensitivity curve. 
We finally extracted the 1D spectra and flux-calibrated them by comparing with a standard $\mathrm{A0V}$ stellar spectrum.
%The latter has been obtained reducing the spectrum of a standard $\mathrm{A0V}$ star as we have done for the scientific frames and dividing the final 1D stellar spectrum by a suitable synthetic model from the \citet{pickles_1998} libraries. 
%We have also applied an airmass correction for the different altitudes at which the standard star and the cluster have been observed ($\sim15-20$\% for the spectral regions of poorest transmission). 
%We have eventually corrected for the flux loss due to the finite width of the slits comparing the integrated flux within the \textit{H} and \textit{Ks} band with the total photometric values. A mean aperture correction factor of $\sim1.3\pm 0.3$ has been applied.\\       
We estimated aperture corrections ($\sim1.3$ on average) comparing the integrated flux within the \textit{H} and \textit{Ks} band with the total photometric values.
As a final step we modelled the noise at each wavelength, taking into account possible slit-to-slit and wavelength-dependent sky variations.\\
We successfully detected ($3\sigma$ confidence level down to an observed \ha\ flux of $1.4\times 10^{-17}\,\mathrm{erg\, cm^{-2}\,s^{-1}}$) at least one line in $71$\% of the sample (78/110 galaxies). In 71, 22, 41, and 7 galaxies we detected at $3\sigma$ \ha, \nii, \oiii, and \hb, respectively. For galaxies where we detected at least one line at $3\sigma$, we put $2\sigma$ upper limits on the other lines, if present in the observed spectral range. When available, we averaged the line fluxes from WFC3 observations with MOIRCS-detected lines, assigning higher weights to higher S/N estimates and properly taking into account the consistency of the \ohb\ ratios and the total flux scaling between the two independent measurements. A total of $49$ galaxies have a detection or a $2\sigma$ upper limit on all \hb, \oiii, \ha, \nii\ emission lines.
%of the four lines necessary to enter the line diagnostic diagrams presented in Section \ref{sec:line_diagnostics}.
\begin{deluxetable}{cccc}
        \tabletypesize{\footnotesize}
        \tablecolumns{10}
        \tablewidth{0pt}
	\tablecaption{Observation log.\label{tab:observation_log}}
        \smallskip
        \tablehead{
          \colhead{Mask ID\tablenotemark{a}} & 
          \colhead{Date\tablenotemark{b}} & 
          \colhead{Integration time\tablenotemark{c}}&
          \colhead{Target field\tablenotemark{d}} \tiny\\
          \colhead{}&
          \colhead{}&
          \colhead{(hr)}& 	
          \colhead{}\\  
        }
	\footnotesize
        \startdata
	\multirow{2}{*}{Mask 1} &	2013 April, $7^{\mathrm{th}}$ &	4.3 &	CL J1449+0856 \\
		               		      &	2013 April, $9^{\mathrm{th}}$ &	   3 &	CL J1449+0856 \\										
                                              &                                                   &        &                                 \\
	\multirow{2}{*}{Mask 2} &	2013 April, $8^{\mathrm{th}}$ &	   5 &	CL J1449+0856 \\
		    				&	2013 April, $9^{\mathrm{th}}$ &	1.7 &	CL J1449+0856 \\										
                                              &                                                   &        &                                 \\
	\multirow{3}{*}{Mask 3} &	2013 April, $7^{\mathrm{th}}$ &	1.7 &	COSMOS \\
		    				&	2013 April, $8^{\mathrm{th}}$ &	1.2 &	COSMOS \\										
		    				&	2013 April, $9^{\mathrm{th}}$ &	0.5 &	COSMOS \\
	\enddata
	\tablenotetext{a}{ID of the three MOIRCS masks.}
	\tablenotetext{b}{Date of observation.}
	\tablenotetext{c}{Total integration time per night.}
	\tablenotetext{d}{Pointed target field.}
\end{deluxetable}
%\begin{table}
%	\caption{Observation log.\label{tab:observation_log}}
%	\smallskip
%	\begin{tabular}{cccc}
%	\toprule
%	\toprule
%	\noalign{\smallskip}
%	Mask ID\tablenotemark{1} &	Date\tablenotemark{2} &	Integration time\tablenotemark{3} &	Target field\tablenotemark{4} \\	
%	& & (hr) & \\			
%	\noalign{\smallskip}
%	\noalign{\smallskip}
%	\midrule
%	\multirow{2}{*}{Mask 1} &	2013 April, $7^{\mathrm{th}}$ &	4.3 &	CL J1449+0856 \\
%		               		      &	2013 April, $9^{\mathrm{th}}$ &	   3 &	CL J1449+0856\\										
%	\multirow{2}{*}{Mask 2} &	2013 April, $8^{\mathrm{th}}$ &	   5 &	CL J1449+0856 \\
%		    				&	2013 April, $9^{\mathrm{th}}$ &	1.7 &	CL J1449+0856 \\										
%	\multirow{3}{*}{Mask 3} &	2013 April, $7^{\mathrm{th}}$ &	1.7 &	COSMOS \\
%		    				&	2013 April, $8^{\mathrm{th}}$ &	1.2 &	COSMOS \\										
%		    				&	2013 April, $9^{\mathrm{th}}$ &	0.5 &	COSMOS \\
%	\bottomrule										
%	%\noalign{\smallskip}
%	\end{tabular}
%	\tablenotetext{1}{ID of the three MOIRCS masks.}\\
%	\tablenotetext{2}{Date of observation.}\\	
%	\tablenotetext{3}{Total integration time per night.}\\
%	\tablenotetext{4}{Pointed target field.}
%\end{table}   

\section{Methodology}
\begin{deluxetable*}{cccccccc}
        \tabletypesize{\footnotesize}
        \tablecolumns{8}
        \tablewidth{\textwidth}
	\tablecaption{Stacked spectra properties. \label{tab:stack_measurements}}
        \smallskip
        \tablehead{
          \colhead{Environment} & 
          \colhead{No. sources} & 
          \colhead{$\langle z \rangle$}&
          \colhead{$\mathrm{log(M)}$}& 
          \colhead{$\mathrm{SFR_{SED}}$} &  
          \colhead{$E(B-V)_{\mathrm{cont}}$} &
          \colhead{$\mathrm{SFR_{H\alpha}}$} &  
          \colhead{$E(B-V)_{\mathrm{neb}}$} \tiny\\
          \colhead{}&
          \colhead{}&
          \colhead{}&
          \colhead{$(\mathrm{log(\msun)})$}& 	
          \colhead{$(\msun\,\mathrm{yr}^{-1})$} & 
          \colhead{} & 
          \colhead{$(\msun\,\mathrm{yr}^{-1})$} & 
          \colhead{}\\  
        }
	\footnotesize
        \startdata
        Cluster& $6$& $1.99$& $10.47$& $101$& $0.29$& $112$&  $0.32$\\
        Field& $31$& $1.92$& $10.57$& $126$&  $0.31$& $68$&   $-$ \\
        Field& $16$& $2.14$& $10.52$& $110$&  $0.33$& $75$& $0.48$\\
        \enddata
       \tablecomments{SED derived quantities are the mean values of single sources in the stacked spectra.}
\end{deluxetable*}
\begin{deluxetable*}{cccccccc}
  %\vspace*{-1cm}
        \tabletypesize{\footnotesize}
        \tablecolumns{8}
        \tablewidth{\textwidth}
	\tablecaption{Stacked spectra observed fluxes. \label{tab:stack_fluxes}}
        \smallskip
        \tablehead{
        \colhead{Environment} & 
        \colhead{No. sources} & 
        \colhead{\oii} &	
        \colhead{\hb} &
        \colhead{\oiii} &	 
        \colhead{\ha} &	
        \colhead{\nii} &
        \colhead{[\ion{S}{2}]$_{\mathrm{tot}}$\tablenotemark{a}} \tiny\\
        \colhead{}&
        \colhead{}&
        \colhead{(cgs)}& 
        \colhead{(cgs)}& 
        \colhead{(cgs)}& 
        \colhead{(cgs)}& 
        \colhead{(cgs)}& 
        \colhead{(cgs)}\\  
        }
	\footnotesize
        \startdata
        Cluster& $6$& $0.800\pm0.065$\tablenotemark{b}&  $0.463\pm0.065$& $1.781\pm0.043$&  $1.915\pm0.061$&  $0.145\pm0.048$&  $0.119\pm0.026$\\
        Field& $31$& $-$& $-$& $-$&  $0.754\pm0.020$&  $0.159\pm0.015$&  $0.075\pm0.011$\\
        Field& $16$& $-$& $0.128\pm0.016$& $0.499\pm0.019$&  $0.645\pm0.016$&  $0.104\pm0.012$&  $0.070\pm0.010$\\
	\enddata
        \tablenotetext{a}{Total combined flux of [\ion{S}{2}]$\lambda\lambda6716,6731$.}
        \tablenotetext{b}{Value for 5/6 cluster members with \oii\ coverage.}
        \tablecomments{The observed line fluxes are expressed in units of $10^{-16}\,\mathrm{erg\,cm^{-2}\,s^{-1}}$. \oii\ fluxes come from WFC3 observations (G13).}
\end{deluxetable*}
\label{sec:methodologies}

\subsection{SED fitting}
\label{sec:SED_fitting}
Stellar masses, SFRs, and dust reddening were determined using FAST \citep{kriek_2009} on the UV to IR photometry. We used \citet{bruzual_2003} models with constant star formation histories (SFHs) and a Salpeter IMF \citep{salpeter_1955}. The \citet{calzetti_2000} reddening law was used to estimate the extinction. %An extra-absorbing component (a ``bump'' at $\sim2200\mathrm{\AA}$) has been taken into account for the field sample in the COSMOS field, where the 30-bands photometry probes this feature. 
For the COSMOS sample, for which the photometric coverage probes the rest-frame UV SED with high accuracy, we allowed for a variable UV bump in the fit \citep{noll_2009}. The slope of the attenuation law was not fitted and the derived SFR estimates are consistent with those derived with the \citet{calzetti_2000} law. We note that the choice of a different SFH, possibly rising or exponentially declining, negligibly affects our mass estimates, well within systematic uncertainties ($\sim0.2\,\mathrm{dex}$). Indeed, for active star-forming galaxies at these redshifts the SED fit gives in most cases very short ages and comparable e-folding times, such that the actual SFH is nearly constant, no matter whether an exponentially increasing or decreasing SFR is used \citep{maraston_2010}. On the contrary, other parameters used in this work are potentially affected by the choice of the SFH, e.g. the SFR. We opted for a constant SFH as it proved to give consistent results in representing the so called ``Main Sequence'' of SFGs \citep[MS,][]{daddi_2007, rodighiero_2014}, matching the SFRs derived independently from \ha\ fluxes and FIR and X-ray stacking. We fitted the photometry for both the aperture-based and the GALFIT-based catalogs (see Section \ref{sec:data_sample} and S13).  Stellar masses and SFRs from the SExtractor catalog were corrected based on total-to-aperture flux ratios ($\lesssim 0.15\,\mathrm{dex}$ for the sample used here). For those galaxies for which the IRAC photometry suffers from a potentially heavy contamination from neighbors ($\sim10$\% of our sample), we excluded the $3.6$-$4.5\mu\mathrm{m}$ bands from the fitting procedure. The two photometric catalogs yield broadly consistent parameter values (e.g., a $\sim0.1\,\mathrm{dex}$ difference in total stellar masses).\\

\subsection{Stacking}  
\label{sec:stacking}
\begin{deluxetable*}{ccccccc}
        \tabletypesize{\footnotesize}
        \tablecolumns{10}
        \tablewidth{\textwidth}
	\tablecaption{Properties of the $6$ confirmed cluster star-forming members in the stack.\label{tab:cluster_sources}}
        \smallskip
        \tablehead{
          \colhead{ID} & 
          \colhead{RA} & 
          \colhead{DEC}&
          \colhead{$z_{\mathrm{spec}}$}&
          \colhead{$\mathrm{log(M)}$}& 
           \colhead{$E(B-V)_{\mathrm{neb}}^{\mathrm{SED}}$\tablenotemark{a}} &
          \colhead{$\mathrm{SFR_{H\alpha}}$}\\   
          \colhead{}&
          \colhead{$(\mathrm{deg})$}&
          \colhead{$(\mathrm{deg})$}&
          \colhead{}&
          \colhead{$(\mathrm{log(\msun)})$}& 	
          \colhead{} & 
          \colhead{$(\msun\,\mathrm{yr}^{-1})$}\\  
        }
	\footnotesize
        \startdata
        ID568& $222.3024796$& $8.9387313$& $1.987\pm0.001$& $10.38$&    $0.33$&      $118\pm9$\tablenotemark{b}\\
        ID510& $222.2997850$& $8.9369198$& $1.988\pm0.001$& $10.52$&    $0.53$&      $158\pm13$\\
        ID422& $222.2983118$& $8.9335256$& $1.988\pm0.001$& $10.53$&    $0.43$&      $361\pm60$\\
        ID183& $222.2961999$& $8.9248673$& $1.990\pm0.001$& $10.05$&    $0.10$&      $24\pm3$\\
        ID580& $222.3070938$& $8.9397864$& $2.001\pm0.001$& $10.54$&    $0.43$&      $189\pm23$\\
        ID41& $222.3029800$& $8.9186500$& $1.991\pm0.001$& $10.63$&    $0.50$&      $125\pm30$\\
	\enddata
        \tablenotetext{a}{$E(B-V)_{\mathrm{neb}}^{\mathrm{SED}}$ is the nebular reddening derived from the SED modelling as $E(B-V)_{\mathrm{cont}}/f$ (Section \ref{sec:ebv}).}
        \tablenotetext{b}{ID568 shows peculiar WFC3 emission line maps \citep{zanella_2015}. Using the emission line maps to compute the aperture correction and the reddening prescription from \citet{zanella_2015} would lead to $\mathrm{SFR_{H\alpha}}=77\pm9\, \msun\,\mathrm{yr}^{-1}$.}
 \end{deluxetable*}

\begin{figure*}
    \includegraphics[width=0.5\textwidth]{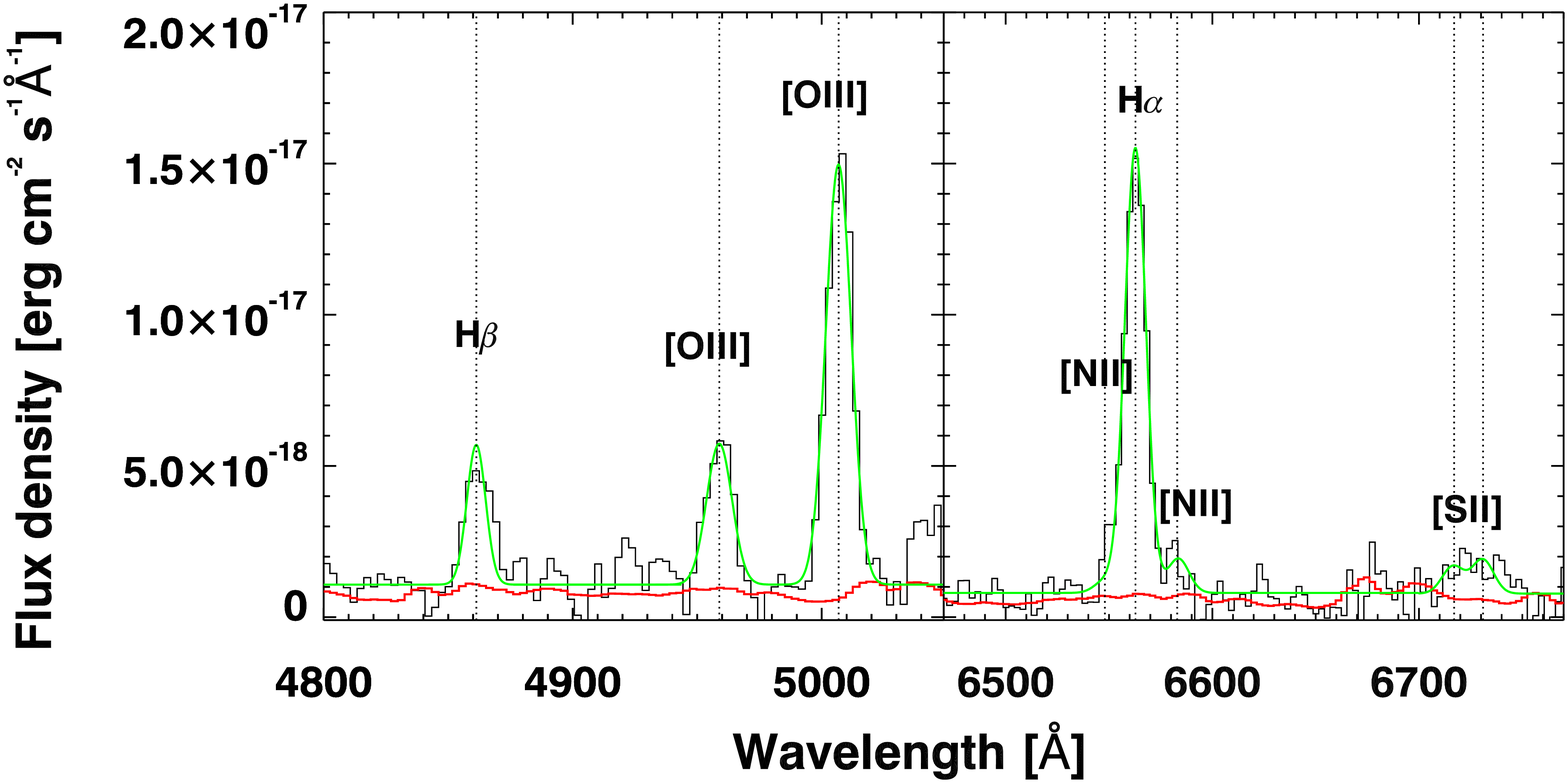}
    \includegraphics[width=0.5\textwidth]{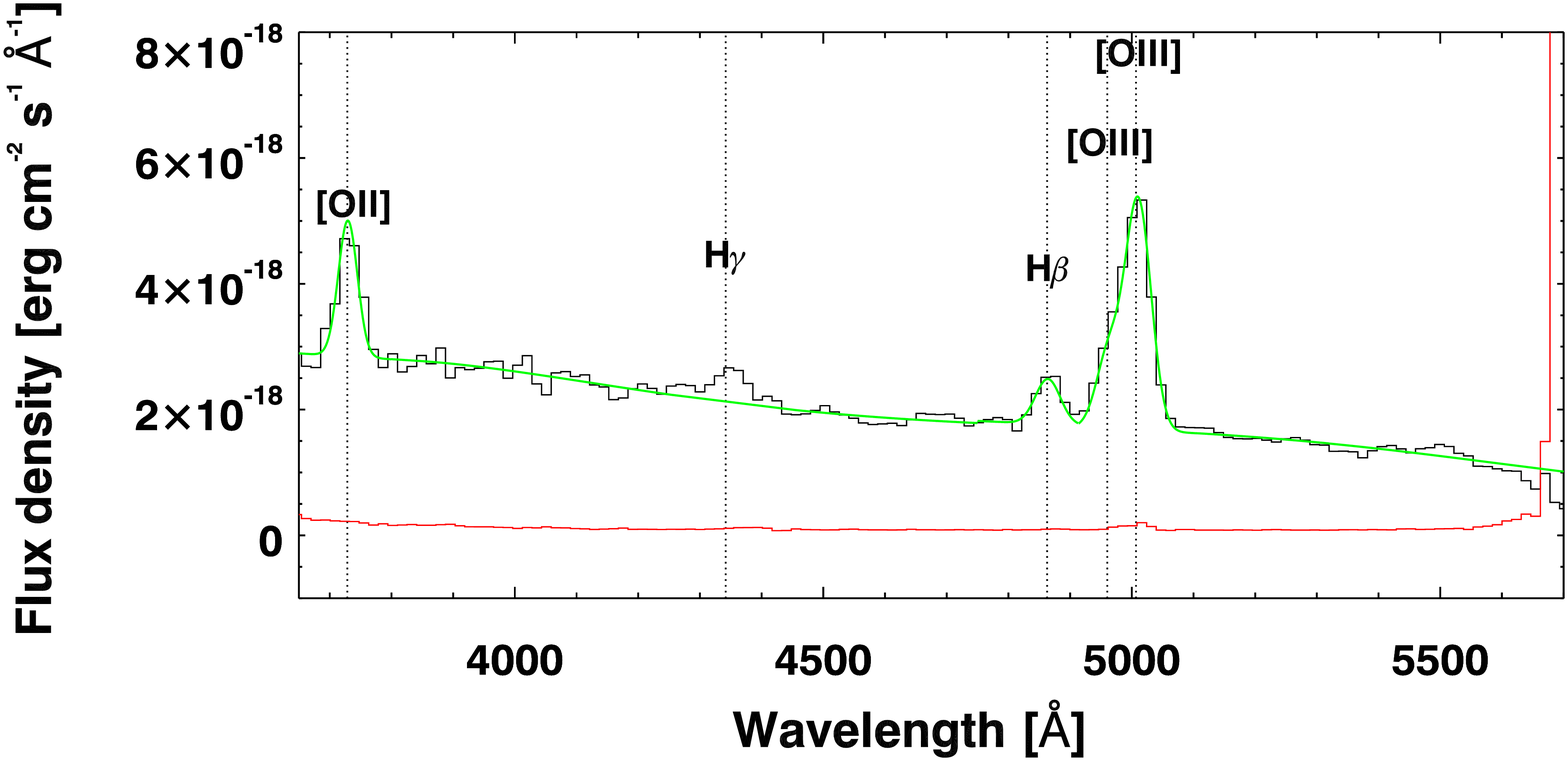}
    \caption{MOIRCS (left panel) and WFC3 (right panel) stacked spectra and noise of the sample of $6$ cluster SFGs in the mass range $10 \leq \mathrm{log(M/\msun)} \leq 11$. The black and red lines respectively represent the stacked spectra and noise. The green line shows the best fit for the emission lines. Vertical dotted lines mark the expected location of emission lines of interest, as labeled.}
    \label{fig:stack_clu}
\end{figure*}
\begin{figure}
    \includegraphics[width=0.5\textwidth]{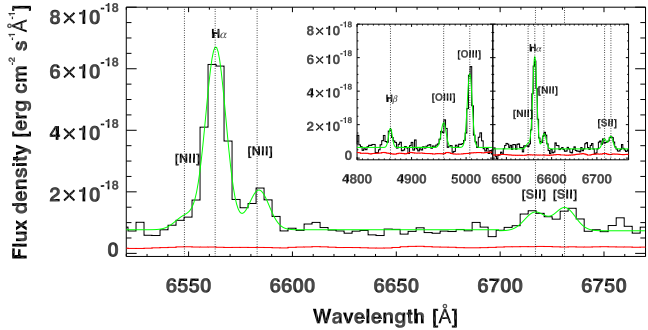}
    \caption{MOIRCS stacked spectrum and noise of the sample of $31$ field SFGs in the mass range $10 \leq \mathrm{log(M/\msun)} \leq 11$ with \ha\ and \nii\ in the observed wavelength range. The black and red lines respectively represent the stacked spectrum and noise. The green line shows the best fit for the emission lines. The onset shows the MOIRCS stacked spectrum and noise for the subsample of $16$ field SFGs in the same mass range with \hb, \oiii, \ha, and \nii\ in the observed wavelength range. Vertical dotted lines mark the expected location of emission lines of interest, as labeled.}
    \label{fig:stack_field}
\end{figure}
\begin{figure}
   \includegraphics[width=0.5\textwidth]{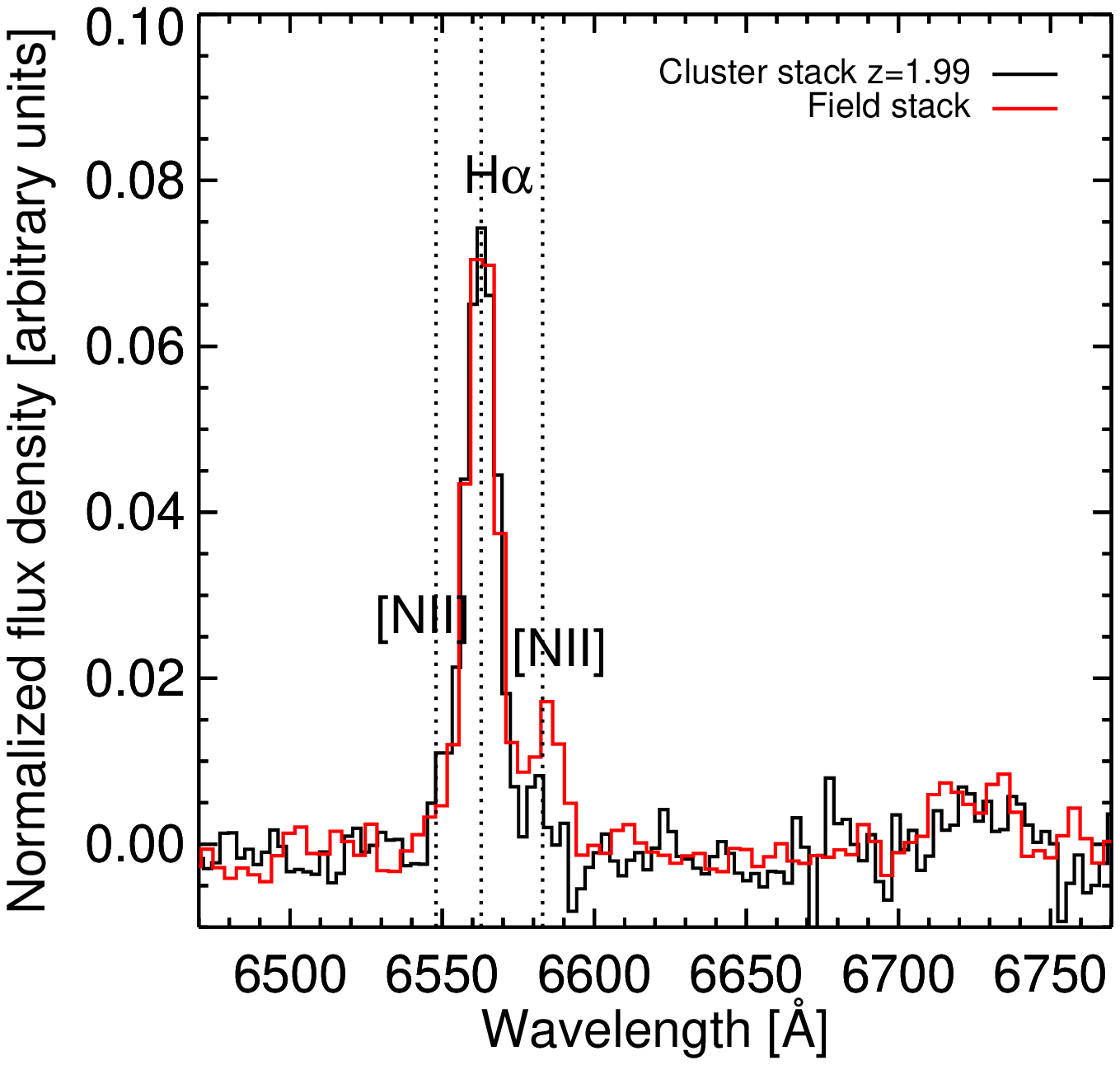}
   \caption{The black and red lines respectively represent the continuum subtracted MOIRCS stacked spectra of the sample of $6$ cluster SFGs and $31$ field SFGs in the mass range $10 \leq \mathrm{log(M/\msun)} \leq 11$ with \ha\ and \nii\ in the observed wavelength range, normalized to \ha\ fluxes. Vertical dotted lines mark the expected location of emission lines of interest, as labeled.}
   \label{fig:stack_normalized}
\end{figure}

In order to maximize the information derivable from the observed spectra and to find an average trend for the cluster and field samples, we stacked individual spectra. We blue-shifted the spectra to the rest-frame and registered them on a common grid of $\sim2.7\,\mathrm{\AA}$ and $\sim3.7\,\mathrm{\AA}$ step for the cluster and field, respectively. Then for every wavelength step we averaged the flux values, weighting for the inverse variance if a sufficiently high number of spectra were co-added ($N>10$). On the other hand, a straight mean was computed in stacking a low number of spectra ($N<10$), not to introduce wavelength dependent biases.
%To preserve the total amount of energy, the flux of each spectrum at any wavelength has been multiplied by $(1+z)$. The same operations have been applied to the noise array. Then for every wavelength step we have averaged the flux values, weighting for $1/N^2(\lambda)$ only if a sufficiently high number of spectra were involved in the stacking ($N>10$). On the contrary, a straight mean has been computed in stacking a low number of spectra ($N<10$) not to introduce wavelength dependent biases. 
We note here that averaging individual spectra does not necessarily coincide with averaging spectral derived quantities. The difference between these two averaged trends depends on the relationship between line fluxes and the derived quantities. In our case we estimated the impact of this difference on the mean metallicity calculated through the line ratio \nha\ for a population of MS--SFGs. For masses $\mathrm{M}\geq 10^{10}\,\msun$, considering the low number statistics for the cluster sample, the two computed averages are similar ($<4$\% difference). Therefore we adopted the mean metallicities coming from the stacking procedure as representative of the population without applying any other correction. The details of this calculation are reported in the Appendix.\\
Considering the low number of sources with $\mathrm{log(M/\msun)} < 10$ and $\mathrm{log(M/\msun)} > 11$, we opted for stacking all galaxies with a spectroscopic redshift determination and \ha\ coverage in the mass range $10 \leq \mathrm{log(M/\msun)} \leq 11$. To investigate possible environmental effects, we stacked the cluster and field sources separately, after a conservative active galactic nuclei (AGN) removal (see Section \ref{sec:line_diagnostics}). We stacked $6$ sources without implementing any weighting scheme for the cluster sample at $z = 1.99$ (Table \ref{tab:cluster_sources}). In the same mass bin we stacked $31$ field sources ($\langle z \rangle = 1.92$) with \ha\ and \nii\ in the observed range and a subsample of $16$ galaxies ($\langle z \rangle = 2.14$) with \hb\, and \oiii\ in addition. Given the number of objects, we applied the optimal weighting described above to the field sample. Unless noted otherwise, in the rest of this work we will use the $31$ sources field stack as a main term of comparison for the analysis of environmental effects to exploit at maximum the sample observed with MOIRCS and we refer to it as the ``field stack''. However, we made use of the $16$-source stacked subsample when necessary, i.e. when \oiii\ or \hb\ fluxes were required. The \nha\ ratio in the 31-source and 16-source field stacks is consistent within the uncertainties. The stacked spectra are shown in Figures \ref{fig:stack_clu} and \ref{fig:stack_field}, and their photometric and spectroscopic properties are summarized in Tables \ref{tab:stack_measurements} and \ref{tab:stack_fluxes}. In Figure \ref{fig:stack_normalized} we plot the continuum subtracted stacked spectra normalized to their \ha\ total fluxes. Furthermore, we checked if the brightest \ha\ emitters biased the average spectra, stacking individual sources normalized by their observed and intrinsic \ha\ fluxes. In both cases \nha\ and \ohb\ are fully compatible with the non- and optimally-weighted measurements within $1\sigma$ error bars. We also stacked only the $5$ cluster sources at a time to check for the impact of low number statistics. In all the cases \nha\ and \ohb\ ratios are consistent within $1\sigma$ with the non-weighted measurements, except for the \ohb\ ratio when stacking only the upper limits on \nii\ (in this case the ratio varies within $2\sigma$ error bars, suggesting possible important physical variance within the sample). We note here that among the $6$ cluster sources in the stack, only the brightest one in \ha\ is detected in \nii\ at $2\sigma$ and corresponds to the lowest \nha\ ratio, which is nevertheless consistent with the average value for the remaining $5$ cluster sources. Therefore, the \nii\ detection is likely the effect of the bright \ha\ emission. Finally, all the $6$ cluster SFGs in the $10 \leq \mathrm{log(M/\msun)} \leq 11$ mass range have a WFC3 spectrum, and for 5/6 sources the $3600 - 5700\,\mathrm{\AA}$ rest-frame interval is covered, giving access to the \oii$\lambda3727$ emission line (\oii\ in the following). Hence, we stacked the WFC3 spectra as done for MOIRCS spectra rescaling to match the absolute fluxes from broad-band photometry. The final stacked \oiii\ and \hb\ fluxes from the WFC3 and MOIRCS spectra result fully compatible within the uncertainties.

\subsection{Line fluxes}
\label{sec:line_fluxes}
We measured line fluxes fitting gaussian profiles to the emission lines on flux calibrated and aperture corrected spectra. We used the squared inverse of the noise array to weight the fitting and to estimate the errors on total fluxes and line positions and thus on the redshift determination. Using the IDL script MPFIT \citep{markwardt_2009}, we fitted at the same time three gaussian profiles lying on a flat continuum to measure [\ion{N}{2}]$\lambda\lambda6548,6583\,\mathrm{\AA}$ and \ha\ fluxes. We modelled the local continuum around each emission line in wavelength ranges large enough to be dominated by continuum emission ($\sim 1000\,\mathrm{\AA}$). In the very few cases where a flat continuum did not provide a good model, we fitted a polynomial curve. We left the \ha\ central wavelength and FWHM free to vary in the fit, while we fixed the \nii\ doublet lines to share a common line width value set by FWHM(\ha) (in terms of velocity), their expected positions relatively to \ha\, and their flux ratio to $3.05$ \citep{storey_2000}. Similarly, we simultaneously fitted the [\ion{O}{3}]$\lambda\lambda4959,5007\,\mathrm{\AA}$ lines, fixing their position and width according to \ha\ values, and their intensity ratio to $2.98$. Any other line in the observed range, both single or in multiplets, was fitted following the same procedure. We estimated flux uncertainties with MPFIT and rescaled them according to the $\chi^2$ value when $\chi^2 > 1.5$. In addition we ran Monte Carlo simulations, placing mock lines in empty spectral regions, recovering consistent uncertainties within $\sim5$\%, confirming the reliability of our noise estimate. We finally estimated the \ha\ and \hb\ stellar absorption measuring the continuum at the proper wavelengths and assuming absorption equivalent widths $\mathrm{EW_{H\alpha}^{abs}} = 3.5\,\mathrm{\AA}$ and $\mathrm{EW_{H\beta}^{abs}} = 5\,\mathrm{\AA}$, as estimated from SED modelling. This correction is $\lesssim 15$\% and $\sim 30$\% for \ha\ and \hb\, respectively.

\subsubsection{Line diagnostics diagrams}
\label{sec:line_diagnostics}
\begin{figure}
\includegraphics[width=0.50\textwidth]{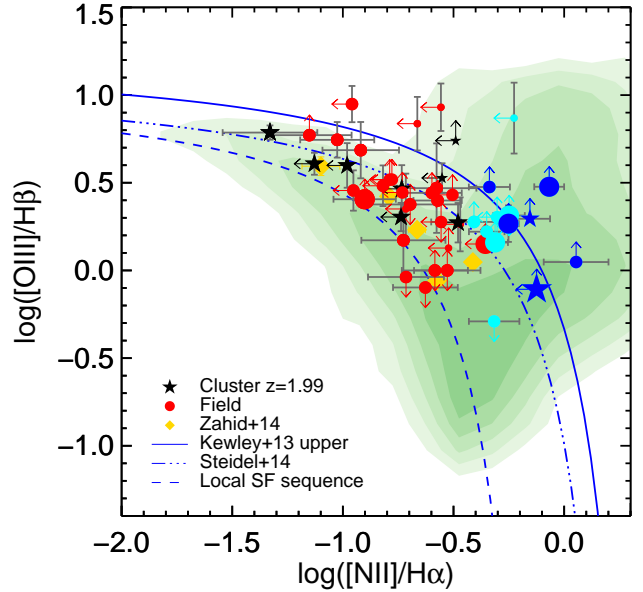}
\caption{BPT diagram for the MOIRCS spectroscopic sample. Red circles and black stars represent the field and cluster samples, respectively. Cyan and blue symbols mark the objects excluded from the SF sample as AGNs from X-ray, radio, \nha\ -- $\mathrm{EW(H\alpha)}$ or the solid curve shown (see text and Figure \ref{fig:ew} for details). Symbol sizes scale as the stellar mass. Golden diamonds represent the stacked points from the FMOS survey at $z\sim1.55$ \citep{zahid_2013}. Arrows indicate $2\sigma$ upper limits both for the $x$ and $y$ axis. The blue dashed line shows the local SF sequence \citep[Eq. 3,][]{kewley_2013p}, the blue dash-dotted line indicates the empirical SF sequence at $z\sim2.3$ from \citet{steidel_2014}, and the blue solid line is the AGN-SFG dividing line at $z=2$ \citep[Eq. 5,][]{kewley_2013p}. Green shaded contours show the SDSS $z\sim 0.1$ sample.}
\label{fig:bpt}
\end{figure}
\begin{figure}
\includegraphics[width=0.50\textwidth]{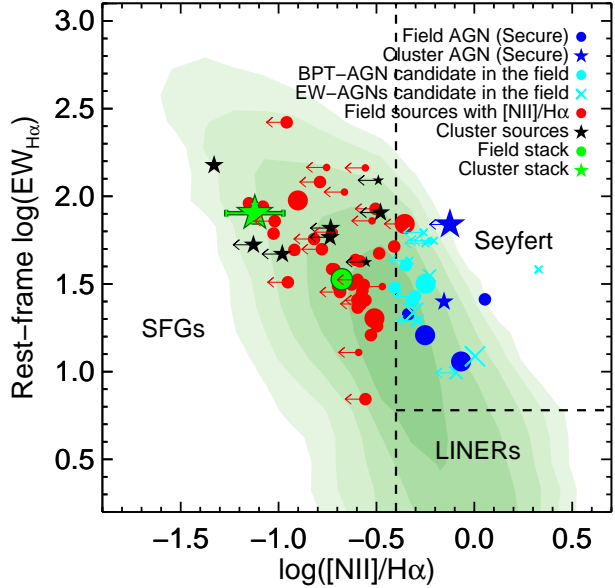}
\caption{\nha\ -- rest-frame reddening uncorrected $\mathrm{EW(H\alpha)}$ diagram for the MOIRCS spectroscopic sample. Red circles and black stars represent the field and cluster samples, respectively. Blue circles and stars respectively represent field and cluster AGNs known from X-ray, radio, and BPT diagram. Cyan symbols represent AGN candidates in the present diagram and in the BPT. Symbol sizes scale as the stellar mass. Arrows indicate $2\sigma$ upper limits. The green circle and star represent the field and cluster stack, respectively. Green shaded contours show the SDSS $z\sim 0.1$ sample.}
\label{fig:ew}
\end{figure}
%\begin{figure}
%\includegraphics[width=0.50\textwidth]{paper_mex_new}
%\caption{MEx diagram for our MOIRCS spectroscopic sample. Red points and black stars represent the field and cluster samples, respectively. Grey symbols mark the objects excluded from the SF sample as AGNs from X-ray, radio, $\mathrm{EW(H\alpha)}$ or the BPT diagram (see text for details). Symbol sizes scale as the stellar mass. Cyan diamonds represent the stacked points from the FMOS survey at $z\sim1.55$ \citep{zahid_2013}. Arrows indicate $2\sigma$ upper limits both for the $x$ and $y$ axis. Solid blue lines are the luminosity dependent tracks proposed by \citet{juneau_2014}, set at the detection limit of our spectroscopic follow-up.}
%\label{fig:mex}
%\end{figure}
% BPT
Following the first pioneering work by \citet{baldwin_1981}, many studies have shown that the proper combination of ratios of collisionally excited and recombination lines can provide useful information not only about the element abundances in the gas in galaxies, but also about its ionization state and the primary ionizing source \citep[e.g.,][]{kewley_2002, kewley_2013p}. In this work we investigate the gas state using the \nha\ -- \ohb\ diagram, commonly referred to as the BPT diagram. This can be used to distinguish line-emitting galaxies mainly powered by an AGN from those dominated by star formation: the radiation field emitted by the disk accreting around an AGN is harder, increasing the oxygen and nitrogen ionization and producing larger \oiii\ and \nii\ fluxes with respect to the values reached by SF-powered ionization. %This results in a clear separation of these two regimes in the \nha\ -- \ohb\ plane. In the local Universe the SFGs form a tight sequence in the BPT diagram, which can be parametrized as the branch of an hyperbola \citep{kewley_2013p} 
%\begin{equation}
%\label{eq:local_sf_sequence}
%\mathrm{ log \left[ \frac{[OIII]}{H\beta} \right]  = \frac{0.61}{log([NII]/H\alpha+0.08)}+1.1}
%\end{equation}      
%with a $\pm0.1\,\mathrm{dex}$ scatter on both \ohb\ and \nha\. A conservative $3\sigma$ upper limit can be used as a ``line of exclusion'' to discern the SF- and the AGN-powered galaxies \citep{kauffmann_2003, kewley_2013p}. 
However the situation may be considerably different at higher redshifts: an evolution of the electron density, the ionization parameter or the hardness of the radiation field can shift the locus of the SF sequence. % and the line of exclusion in other loci of the BPT diagram. 
Recent developments in multi-object near-IR spectroscopy have allowed for the observation of an increasing number of samples of line emitters, extending the study of the potential evolution of line ratios with cosmic time \citep[e.g.,][]{kewley_2013p, holden_2014} and the role of selection effects \citep{juneau_2014} to earlier epochs, up to $z\geq1.5$. Recent results by \citet[][S14 in the following]{steidel_2014} for a sample of $z\sim2.3$ SFGs point towards a substantial vertical shift in the BPT diagram due to a harder field ionizing the ISM, qualitatively in agreement with some theoretical expectations \citep[][but see, e.g., \citealt{coil_2014} for alternative results]{kewley_2013p, kewley_2013l}. Interestingly, S14 interpret the locus of SFGs mainly as a ``ionization parameter'' sequence, in contrast to the usual interpretation of a ``gas-phase metallicity'' sequence given in the local Universe \citep{kewley_2008}. %This links the SFGs position in the BPT diagram to some physical quantitities more fundamental than the gas-phase metallicity. 
Figure \ref{fig:bpt} shows our sample in the BPT diagram and a reference sample at $z\sim 1.55$ from the Subaru/FMOS survey \citep{zahid_2013}. A low-redshift ($0.04<z<0.2$) sample of 299,098 galaxies selected from SDSS DR7 \citep{abazajian_2009} is shown for comparison. Following \citet{juneau_2014}, galaxies were selected to have well constrained \ohb\ and \nha\ line ratios (S/N$>3/\sqrt{2}$, corresponding to each line with S/N$>3$ or to combinations of a weaker and a stronger line, provided that the overall line ratio is constrained to this minimum significance).  Line flux measurements and uncertainties were taken from the MPA/JHU catalogs, and adjusted as detailed by \citet{juneau_2014}. A systematic shift with respect to the locus of SFGs in the local Universe is present, qualitatively in agreement with a possible increase of the hardness of the radiation field, even if the data at our disposal do not allow to recognize a specific direction of the shift. To exclude AGNs from our sample, we used the conservative line of exclusion as a function of redshift provided by Equation 5 in \citet{kewley_2013p}. Alternative emission line diagnostics relying on \ohb\ and either host color or stellar mass have been developed \citep[e.g.,][]{yan_2011,juneau_2011}. However $\sim50$\% of the field stacked sample ($15/31$ sources) does not have \hb\ in the observed range. To obviate this issue, we coupled the BPT diagnostics with the \nha\ -- observed \ha\ equivalent width ($\mathrm{EW(H\alpha)}$) diagram (Figure \ref{fig:ew}). The local SDSS sample is shown again for comparison (here we considered only galaxies with S/N$(\mathrm{EW(H\alpha)}) > 3$, cutting the BPT local sample to 272,562 objects). In this diagram, \nha\ traces the ionized gas conditions, as higher \nha\ values are connected to harder powering sources, while $\mathrm{EW(H\alpha)}$ measures the power of the ionizing source in relation with the continuum emission of the underlying stellar populations \citep{cid-fernandes_2010, cid-fernandes_2011}. In this diagram, all the potential AGNs that we selected on the BPT basis occupy the same region at high \nha\ ratios.  
In total, we conservatively excluded $22$ objects as AGN-powered sources and none of these sources was included in the stacked spectra. Four points above the nominal line of exclusion in the BPT were not discarded as their upper limits on \nha\ are still compatible with the star-forming region in the \nha\ -- observed $\mathrm{EW(H\alpha)}$ diagram. We note that $3/4$ objects have $\mathrm{log(M/M_{\odot})} < 10$ and thus are not part of the stacked spectra. Excluding the fourth BPT potential outlier from the field stacked spectrum would slightly strengthen the significance (well within the uncertainties) of the main results of this work, increasing the \nha\ field average value (see below). Even if we cannot exclude potential AGN contamination for these sources, we lack definitive evidence that they are mainly AGN--dominated and, keeping the most conservative approach in terms of significance of the final results, we retained these four objects in the final samples.
Among these $22$ sources, $2$ are known to be a soft and a hard X-ray AGN in CL J1449+0856, both with $\mathrm{log(M/M_{\odot})} > 11$ (G13), while other $2$ field objects are massive radiogalaxies in the COSMOS field. All these independently known AGNs lie either above the line of exclusion in the BPT diagram or above $\mathrm{log}($\nha$) = -0.4$ in the \nha\ -- observed $\mathrm{EW(H\alpha)}$ plane, as expected. We note here that the choice of the AGNs to remove does not change if we consider a dereddened $\mathrm{EW(H\alpha)}$.
In Figure \ref{fig:ew} we show in addition the position of the cluster and field stacked values. Comparing these two, we note that the cluster and field samples show a $>4\sigma$ significant difference in \nha. Considering the subsample of $16$ field galaxies with all the BPT lines, the difference is still tentatively present ($\sim2.7\sigma$), even if the significance is reduced due to lower number statistics and signal-to-noise. Moreover, in this case the cluster and field \ohb\ ratios are fully compatible within the error bars ($0.585 \pm 0.062$ and $0.591 \pm 0.058\,\mathrm{dex}$, respectively). As a consequence, the (\oiii$/$\hb$)/($\nii$/$\ha) ratio is compatible between the two samples, given the increased uncertainties. Figure \ref{fig:ew} shows also a $0.37\,\mathrm{dex}$ difference ($\sim4.7\sigma$ significant) in the observed $\mathrm{EW(H\alpha)}$ between the cluster and the field, which reflects the $2.5\times$ higher observed \ha\ luminosity in the cluster stack (see Section \ref{sec:ew_ha} for further discussion). 
Finally, we observe a significant [\ion{S}{2}]$\lambda\lambda6716,6731$ emission in the stacked spectra, but the S/N is not high enough to accurately measure the ratio of the two lines and hence directly estimate the electron density $n_{\mathrm{e}}$ \citep{osterbrock_2006}. Therefore, we fixed this ratio compatibly with typical $n_{\mathrm{e}}$ values in [\ion{H}{2}] regions ($n_{\mathrm{e}} = 100-1000\,\mathrm{cm}^{-3}$, \citealt{osterbrock_2006}) and measured the total combined flux [\ion{S}{2}]$_{\mathrm{tot}}=$[\ion{S}{2}]$\lambda6716+\lambda6731$ reported in Table \ref{tab:stack_fluxes}.

\subsubsection{Gas-phase metallicities}
\begin{deluxetable*}{cccccc}
        \tabletypesize{\footnotesize}
        \tablecolumns{6}
        \tablewidth{\textwidth}
	\tablecaption{Gas-phase metallicity estimates for the stacked spectra. \label{tab:stack_metallicities}}
        \smallskip
        \tablehead{
          \colhead{Environment} & 
          \colhead{$\mathrm{log(M)}$} & 
          \multicolumn{4}{c}{$12+\mathrm{log(O/H)}$\tablenotemark{a}} \vspace{0.05cm} \tiny\\  
          \colhead{} &
          \colhead{}\vspace{0.05cm} &
          \multicolumn{2}{c}{$N2$}\vspace{0.05cm}&%=\mathrm{log}($\nii$/$\ha$)$}\vspace{0.05cm}  &
          \multicolumn{2}{c}{$O3N2$}\vspace{0.05cm}\\%=\mathrm{log}\{$\oiii$/$\hb$)/($\nii$/$\ha$)\}$}\vspace{0.05cm}\\
          \colhead{}&
          \colhead{$\mathrm{log(\msun)}$}& 	
          \colhead{PP04\tablenotemark{b}} & 
          \colhead{S14\tablenotemark{c}} & 
          \colhead{PP04\tablenotemark{b}}& 
          \colhead{S14\tablenotemark{c}}\\
        }
	\footnotesize
        \startdata
        Cluster& $10.47$& $8.261 \pm 0.083$&  $8.216 \pm 0.053$& $8.184 \pm 0.051$&  $8.182 \pm 0.044$\\
        Field& $10.57$& $8.514 \pm 0.025$& $8.376 \pm 0.016$&  $8.287 \pm 0.025$\tablenotemark{d}&  $8.273 \pm 0.022$\tablenotemark{d}\\
	\enddata
        \tablenotetext{a}{For comparison, the solar value is $12+\mathrm{log(O/H)}=8.69$ \citep{asplund_2009}.}
        \tablenotetext{b}{\citet{pettini_2004} calibration.}
        \tablenotetext{c}{\citet{steidel_2014} calibration.}
        \tablenotetext{d}{Values for the subsample of $16$ field SFGs with \hb\ and \oiii\ in the observed wavelength range.}
        %\tablecomments{All the errors in the table do not take into account the uncertainties on the metallicity calibration.}
\end{deluxetable*}   
\begin{figure}
\includegraphics[width=0.50\textwidth]{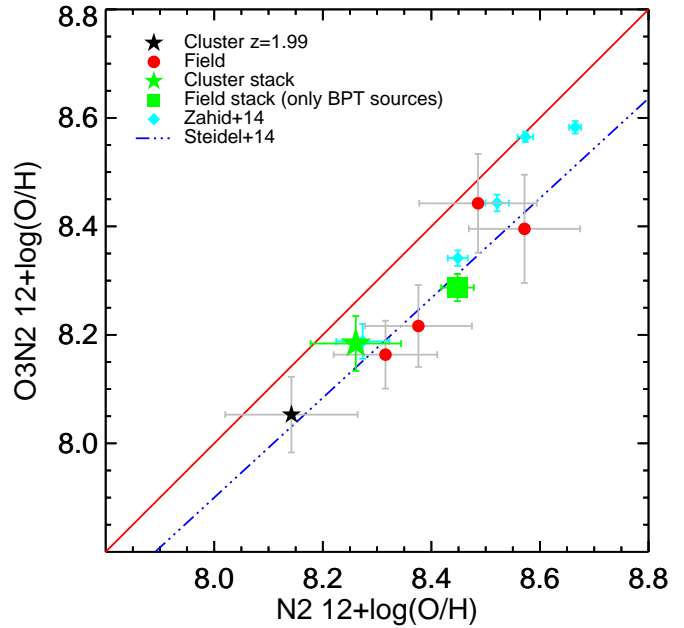}
\caption{Gas-phase metallicities derived using the $N2$ and $O3N2$ indicators calibrated by \citet{pettini_2004}. Red circles and black stars respectively mark individual field and cluster $3\sigma$ detections of each line in the $O3N2$ indicator. The green filled circle and star respectively represent the measurement for the subsample of $16$ field SFGs and cluster stacked sample in the $10 \leq \mathrm{log (M/\msun)} \leq 11$ range. Cyan circles mark the stacked values at $z\sim1.55$ from \citep{zahid_2013}. The blue dash-dotted line is the linear relation between $N2$ and $O3N2$ for the $z\sim2.3$ from \citet{steidel_2014}. A one-to-one red line is shown as a comparison.}
\label{fig:met_indicators}
\end{figure}
Different methods have been proposed through the years to estimate the gas-phase metallicity in galaxies. The safest method involves the ratio of the \oiii$\lambda4363$ auroral line and lower excitation lines as \oiii$\lambda\lambda4959,5007$, which allows to directly evaluate the oxygen abundance through the gas electron temperature ($T_e$). However, \oiii$\lambda4363$ is weak even in low-metallicity regions and generally difficult to measure in high redshift galaxies. Other empirical methods have been proposed to circumvent this problem, calibrating the ratios of stronger lines against $T_e$ in \ion{H}{2} regions. Alternatively, theoretical photoionization models may be employed to predict the line fluxes and derive the gas-phase abundances \citep[see][for a census of gas-phase metallicity calibrations]{kewley_2008}. In general, the use of different methods leads to a systematic difference in absolute metallicity values of up to $\sim0.3\,\mathrm{dex}$ \citep{kewley_2008}. Relative comparisons among different samples from different studies are still meaningful if all the measurements are reported to the same calibration system. For this work we decided to use the $N2=\mathrm{log}($\nii$/$\ha$)$ metallicity indicator, given the presence of both \nii\ and \ha\ in a relatively clear window of the $K_s$ band at $z=2$. \citet[][PP04 in the following]{pettini_2004} calibrated $N2$ against the $T_e$ method in a local sample of \ion{H}{2} regions, expressing the gas-phase metallicity as:
\begin{equation}
12+\mathrm{log(O/H)_{N2,PP04}} = 0.57\times N2+8.90
\end{equation}           
with a quoted uncertainty of $\sim 0.18\,\mathrm{dex}$. Partial drawbacks of using $N2$ are its sensitivity to the ionization parameter $\mathcal{U}$ and the saturation of the index at solar metallicities and above, as \nii\ becomes the dominant coolant \citep{baldwin_1981, kewley_2002}. The impact of this saturation seems not to dramatically affect the final metallicity estimate, resulting in a $\sim0.03\,\mathrm{dex}$ underestimate of the final abundance \citep{zahid_2014}. Other indicators do not suffer from this saturation issue and could potentially be used to confirm the metallicity estimate. When all lines were available, we used the $O3N2=\mathrm{log}[($\oiii$/$\hb$)/($\nii$/$\ha$)]$ as an alternative metallicity indicator. In this case the PP04 calibration gives:
\begin{equation}
12+\mathrm{log(O/H) _{O3N2,PP04}} = -0.32\times O3N2+8.73
\end{equation}           
with a quoted uncertainty of $\sim 0.14\,\mathrm{dex}$. The inclusion of \oiii\ in the ratio should guarantee sensitivity to increasing metallicity even above solar, as \oiii\ continues to decrease while \nii\ saturates. In practice we could estimate this index at $3\sigma$ confidence only for very few individual sources and for the stacked spectra (Figure \ref{fig:met_indicators}). In the local Universe these two indicators provide consistent metallicity estimates \citep{kewley_2008}. On the contrary, for our samples of high-z galaxies the $N2$ indicator returns systematically higher gas-phase metallicities compared to the $O3N2$ indicator (Figure \ref{fig:met_indicators}), in agreement with other recent findings \citep[S14]{erb_2006, yabe_2012, zahid_2013}. Different interpretations and prescriptions to avoid systematic errors have been proposed in several studies (i.e., S14), even if all of them remain quite speculative in absence of a direct $12+\mathrm{log(O/H)}$ measurement, i.e. by means of the $T_e$ method. However, all the studies agree on a probable overall change of the ISM conditions in high-z galaxies with respect to the local Universe, as indicated by independent observations \citep{magdis_2012, kashino_2013}. In principle an evolution in the hardness of the radiation field, electron density, ionization parameter, or nitrogen-to-oxygen ratio can make the calibration intrinsically wrong for high-redshift galaxies. In their recent work, S14 recalibrated the $N2$ indicator on a sample of local \ion{H}{2} regions matching the physical conditions of their $z\sim 2.3$ galaxies, obtaining:
\begin{equation}
12+\mathrm{log(O/H)_{N2,S14}} = 0.36\times N2+8.62
\end{equation} 
with a quoted total scatter of $\sim 0.13\,\mathrm{dex}$. S14 found that $N2$ is less sensitive to metallicity variations than implied by the PP04 calibration, which substantially overpredicts the metallicities at high redshifts. On the contrary, after S14 recalibration, the $O3N2$ indicator predicts metallicities similar to those given by the PP04 calibration, especially with the inclusion of a term depending on N/O \citep{perez-montero_2009}:
\begin{equation}
12+\mathrm{log(O/H) _{O3N2,S14}} = -0.28\times O3N2+8.66
\end{equation}
with a total uncertainty of $\sim 0.12\,\mathrm{dex}$. Reducing the sensitivity of the $N2$ calibrator to the gas-phase metallicity and leaving intact the one of $O3N2$, the two ratios predict consistent abundances at $z\sim2$.\\
%All the gas-phase metallicity estimates are reported in Table \ref{tab:stack_metallicities}: \textit{in the same mass range, the cluster sample results always more metal poor than the field sample at face value, irrespectively of the calibration used}. On the other hand, \textit{the amount of metal deficiency and the significance of the cluster-to-field difference depends on the sensitivity to O/H assigned to the indicators} (see Section \ref{sec:mzr} for a further discussion).\\
As mentioned, a cause of concern when estimating gas-phase oxygen abundance through indirect indicators involving other species as $N2$ and $O3N2$ is the abundance of these elements relative to oxygen. In the case of $N2$ and $O3N2$ indicators, an assumption on the N/O ratio is implied in every calibration, explicitly or implicitly, and ignoring the N/O ratio could result in a systematic effect in the O/H estimation \citep{perez-montero_2009}. An estimation of N/O can be derived from the $N2O2 =\mathrm{log}($\nii$/$\oii$)$ ratio, as calibrated in the local Universe by \citet{perez-montero_2009}:
\begin{equation}
\mathrm{log(N/O)} = 0.93\times N2O2-0.20
\end{equation}
with a standard deviation of the residuals of $0.24\,\mathrm{dex}$. We could estimate N/O for the cluster stacked sample thanks to the WFC3 \oii\ determination after proper dust reddening correction (see next Section). This was not possible for the field sample, preventing a fully consistent environmental comparison of N/O. In our case, the inclusion of a N/O correction term in the PP04 $N2$ calibration \citep[Equation (13) from][]{perez-montero_2009} leaves virtually unchanged the metallicity estimate for the cluster (a $\sim0.05\,\mathrm{dex}$ difference, well within the calibration errors). The observed cluster N/O ratio ($\mathrm{log(N/O)} = -1.18\pm0.15$) is lower than the solar value \citep[$\mathrm{log(N/O)} \simeq -0.86$,][]{pilyugin_2012}, and close to the ``primary'' nitrogen abundance predicted by current models ($\mathrm{log(N/O)} \simeq -1.5$, \citealp{charlot_2001, perez-montero_2009, pilyugin_2012, andrews_2013, dopita_2013}) in agreement with the estimated low gas-phase metallicity value. Interestingly, recent works on samples at redshift $z\sim 2$ found a N/O ratio consistent with the solar value and only slowly or not varying with the O/H ratio, and hence with the gas-phase metallicity, at least for highly star-forming systems typical at these redshifts \citep[$\mathrm{SFR} \geq 10\,\msun\,\mathrm{yr^{-1}} $,][S14]{andrews_2013}. An alternative explanation is that $z\sim2$ SFGs show higher N/O ratios at fixed metallicities than local counterparts at low masses ($\mathrm{M\lesssim10^{10.11}\,\msun}$ in \citealt{shapley_2014}, and $\mathrm{M\sim10^{9}\,\msun}$ \citealt{masters_2014}). If we consider N/O ratios from literature as representative of a general field sample (but check \citealp{kulas_2013} for the study of protocluster members in S14 sample) in a mass range and excitation conditions similar to those of our cluster sample, they result to be $\sim0.2\,\mathrm{dex}$ higher than the value measured on the cluster stacked spectrum.\\ 
Further indications of the lower metal content in cluster sources are the lower \nii$/$[\ion{S}{2}]$_{\mathrm{tot}}$ and higher $($\ha$+$\nii$)/$[\ion{S}{2}]$_{\mathrm{tot}}$ ratios than the field counterparts \citep{nagao_2006}, even if affected by substantial uncertainties. Moreover, for the cluster stacked sample we estimated the ionization parameter and the gas-phase metallicity using the iterative method of \citet{kobulnicky_2004}, which involves the \oii, \hb, and \oiii\ fluxes. After applying the \citet{kewley_2008} conversion to PP04 $N2$ metallicities, we obtain $12+\mathrm{log(O/H)} = 8.217$ with $\sim0.15\,\mathrm{dex}$ accuracy, thus compatible with our estimate based on \nha\ (Table \ref{tab:stack_metallicities}). We also obtained a ionization parameter $\mathcal{U}\simeq-2.61$, which is comparable to values measured in high redshift galaxies \citep[$-2.9<\mathcal{U}<-2.0$,][and references therein]{kewley_2013p}. Hence a high ionization parameter may not be the main driver of the \nha\ difference that we observed between cluster and field, even if we cannot completely exclude possible effects connected to $\mathcal{U}$ in our analysis.\\
Overall, while various metallicity estimators may differ on an absolute scale, the systematic difference found between the cluster and the field is robust. Given the lower number of sources with a safe $O3N2$ measurements, we privileged $N2$ as primary metallicity indicator. All the gas-phase metallicity estimates are reported in Table \ref{tab:stack_metallicities}: in the same mass range, the cluster sample results $0.09-0.25\,\mathrm{dex}$ (using $O3N2_{\mathrm{S14}}$ and $N2_{\mathrm{PP04}}$, respectively) more metal poor than the field sample, depending on the calibration used.           

\subsubsection{Nebular E(B-V) estimate}
\label{sec:ebv}
\begin{figure}
  \includegraphics[width=0.50\textwidth]{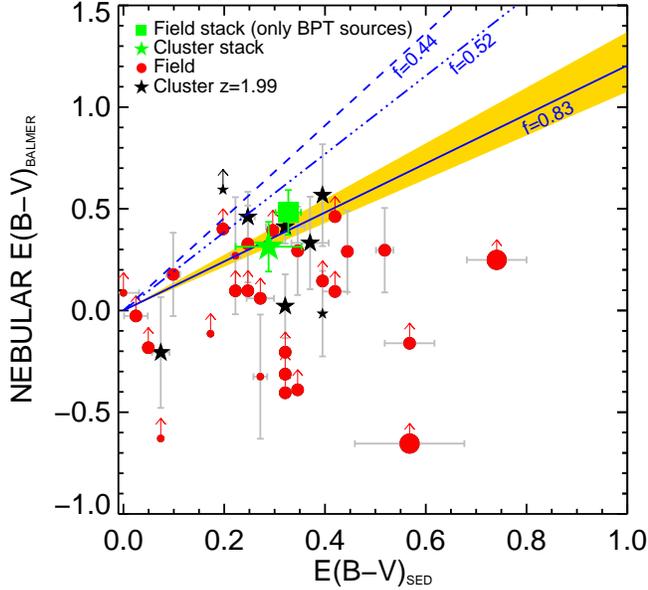}
  \caption{Reddening estimates based on SED fitting and Balmer decrement. Red circles and stars respectively mark field and cluster SFGs with $3\sigma$ \ha\ and \hb\ detections. Symbol sizes scale as stellar mass. Arrows mark $3\sigma$ lower limits. The green square and star indicate the subsample of $16$ field sources with measured \oiii\ and \hb\ and the cluster stacked values, respectively. The blue dashed and dotted-dashed lines represent the $f=0.44$ ($f=0.52$) ratio $E(B-V)_{\mathrm{cont}}/ E(B-V)_{\mathrm{neb}}$ obtained in the local Universe applying Fitzpatrick-Calzetti (Calzetti-Calzetti) laws for the nebular and continuum reddening, respectively. The blue solid line and the shaded area represent the same ratio using a Calzetti law for both the nebular and continuum reddening and the relative uncertainties quoted in \citet{kashino_2013} for the sample of $z\sim1.55$ galaxies from the FMOS survey, where $f=0.83$.}
  \label{fig:ebv_kashino}
\end{figure}
\begin{figure}
  \includegraphics[width=0.50\textwidth]{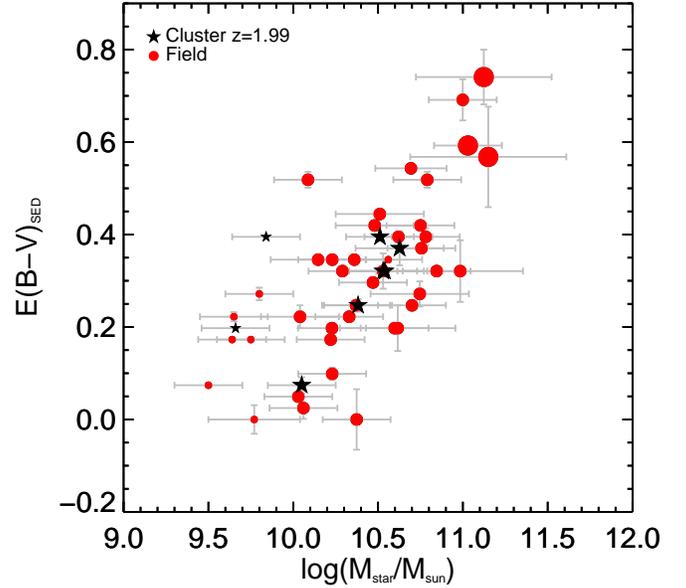}
  \caption{Mass-Reddening Relation for the MOIRCS spectroscopic sample of SFGs. Red circles and black stars mark the field and cluster samples, respectively. Symbol sizes scale as the stellar mass.}
  \label{fig:mrr}
\end{figure}
The dust reddening on stellar light ($E(B-V)_{\mathrm{cont}}$) was estimated through SED fitting. However, the amount of dust attenuation toward the emission lines ($E(B-V)_{\mathrm{neb}}$) is typically larger than $E(B-V)_{\mathrm{cont}}$. \citet{calzetti_2000} find a factor $f=E(B-V)_{\mathrm{cont}}/ E(B-V)_{\mathrm{neb}} = 0.44$ between the two color excesses in the local Universe, adopting the \citet{fitzpatrick_1999} law for the nebular reddening and their own law for the continuum reddening ($f=0.52$ using the \citet{calzetti_2000} reddening law for both the nebular emissions and the continuum). Recent works suggest that this continuum-to-nebular emission differential reddening factor is generally higher for high-redshift galaxies, reducing the difference between stellar and nebular continuum \citep{kashino_2013,pannella_2014}. Here we attempt to estimate this factor using the Balmer decrement $\mathrm{H\alpha/H\beta}$ and assuming a Case B recombination with a gas temperature $T =10^4\,\mathrm{K}$ and an electron density $n_e = 100\,\mathrm{cm^{-3}}$, according to which the intrinsic ratio $\mathrm{H\alpha/H\beta}$ is equal to $2.86$ \citep{osterbrock_2006}. From the observed \ha\ and \hb\ values it is possible to compute:
\begin{equation}
E(B-V)_{\mathrm{neb}} = \frac{2.5}{k_{\mathrm{H\beta}} - k_{\mathrm{H\alpha}}} \mathrm{log \left[ \frac{H\alpha/H\beta}{2.86} \right]}
\end{equation}
assuming a proper extiction law. In this work we assumed the \citet{calzetti_2000} law for both the nebular and continuum reddening, for which $k_{\mathrm{H\beta}}=4.598$ and $k_{\mathrm{H\alpha}}=3.325$.\\
A limited sample of galaxies with a safe $3\sigma$ \hb\ detection is available to measure the Balmer decrement on an object-by-object basis. However, we used the values from the stacked spectra to assess this issue for the mean population of SFGs in our sample. In Figure \ref{fig:ebv_kashino} we show the relation between SED based $E(B-V)_{\mathrm{cont}}$ and $E(B-V)_{\mathrm{neb}}$ derived from the Balmer decrement. %Despite the larger scatter and the low number of significant $\mathrm{H\beta}$ $3\sigma$ detections, the points are preferentially located around the $f=0.83$ line, pointing towards a smaller difference between the stellar and the nebular reddening at $z\sim2$. This result is visually clearer if we consider only the stacked values in the intermediate mass regime. 
The best fitting slope for the $3\sigma$ detected stacked values is $0.74\pm0.22$, consistent within the uncertainties with the results from the FMOS survey at $z\sim1.55$ \citep{kashino_2013}, but still formally in agreement with the local value obtained using the same reddening law. Our best fit value is in agreement with the alternative estimate that we derived from the fitting of the $E(B-V)_{\mathrm{cont}}$-$\mathrm{SFR_{H\alpha}^{uncorr}/SFR_{UV}^{uncorr}}$ relation as in Figure 3 of \citet{kashino_2013}, namely $f=0.74\pm0.05$.\\
We checked for possible environmental signatures in the stellar Mass-Reddening Relation (MRR) comparing the cluster and field samples. Figure \ref{fig:mrr} shows the MRR for our sample of cluster and field SFGs. Both the stellar mass and the reddening estimates come from the SED fitting procedure. In Figure \ref{fig:mrr} both the cluster and field samples follow the same trend, not revealing any environmental signature in the MRR. Applying a simple linear regression separately for the two samples we obtain compatible slopes: $0.78\pm0.42$ and $0.59\pm0.07$  for the cluster and field sample, respectively.

\subsubsection{A significant difference in the observed $\mathrm{EW(H\alpha)}$}
\label{sec:ew_ha}
\begin{figure*}
  \includegraphics[width=\textwidth]{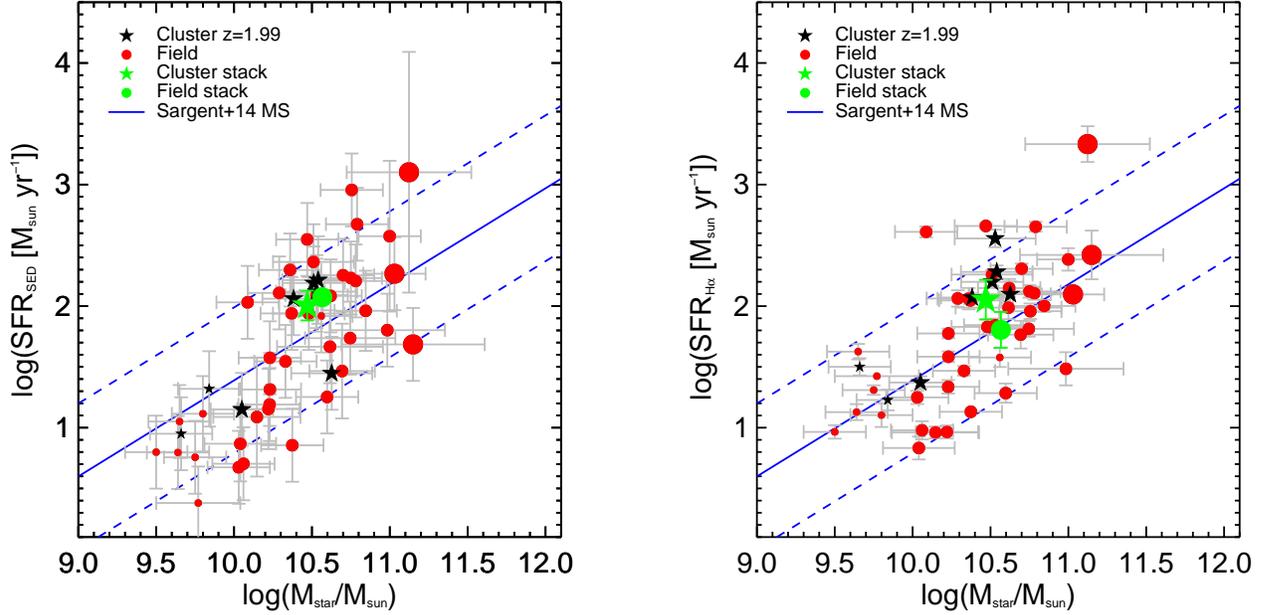}
  \caption{Stellar mass versus SFR. Red circles and black stars mark the field and cluster SFGs, respectively. Symbol sizes scale as the stellar mass. The green circle and star indicate the field and cluster stacked values, respectively. The blue solid line indicates the MS at $z=2$ as parametrized in \citet{sargent_2014}. The blue dashed lines mark the $\pm0.6\,\mathrm{dex}$ scatter. Left panel: \ha\ derived SFR. Right panel: SED derived SFR.}
  \label{fig:ms}
\end{figure*}
\begin{figure}
  \includegraphics[width=0.50\textwidth]{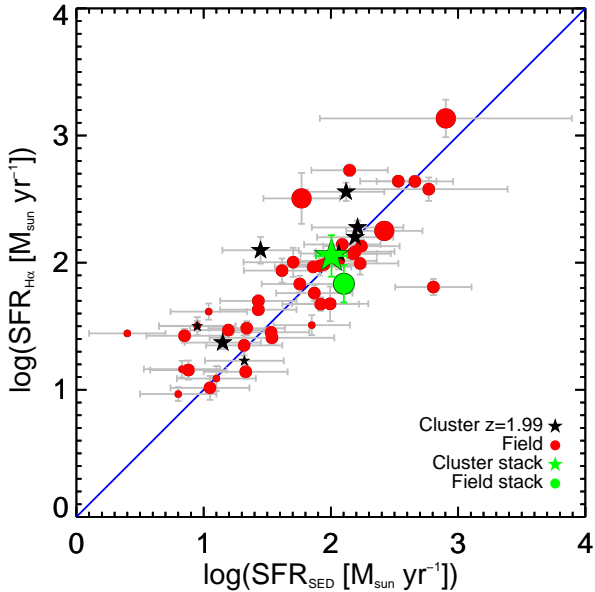}
  \caption{SFR estimates from SED fitting and intrinsic \ha\ luminosities for the MOIRCS spectroscopic sample of SFGs. Red circles and black stars respectively mark the field and cluster samples with $2\sigma$ detected \ha\ line. Symbol sizes scale as the stellar mass. The green circle indicates the field stacked value. The green star indicates the cluster stacked value. A one-to-one blue line is shown as a comparison.}
  \label{fig:sfr}
\end{figure}
 As shown in Figure \ref{fig:ew}, there is a $0.37\,\mathrm{dex}$ difference ($\sim4.7\sigma$ significant) in the observed $\mathrm{EW(H\alpha)}$ between the cluster and the field. Such a difference may arise from an enhanced sSFR, a variation in the dust reddening correction $E(B-V)$, or in the continuum-to-nebular emission differential reddening factor $f$ between cluster and field, as $\mathrm{EW(H\alpha)} \propto \mathrm{sSFR}\times10^{0.4E(B-V)_{\mathrm{cont}}k_{\mathrm{H}\alpha}(1/f-1)}$. Assuming a common $f$ value in cluster and field SFGs and the average $E(B-V)_{\mathrm{cont}}$ values in Table \ref{tab:stack_measurements}, the difference in the observed $\mathrm{EW(H\alpha)}$ is translated into a significant difference in intrinsic $\mathrm{EW(H\alpha)}$ and ascribable to enhanced sSFR in cluster sources. As $f$ is physically linked to the average geometric dust distribution in galaxies’ star-forming regions \citep{kashino_2013}, there are not immediately evident reasons why the environment should play a role in setting this factor. Hence, considering $f$ constant within different environments would not be a strong assumption. However, we could let this parameter free as well, resulting in a more conservative approach: in this case, the $f$ factor for the field stack is tentatively lower than for the cluster sample, reducing the difference in intrinsic $\mathrm{EW(H\alpha)}$. Moreover, individual estimates of $f$ are hampered by large error bars on the Balmer decrement measurements (Figure \ref{fig:ebv_kashino}), not allowing to fully decouple sSFR and reddening effects. Since the two stacked samples have similar stellar masses, an enhancement in sSFR would reflect the $2.5\times$ higher \ha\ observed luminosity in the cluster stack (Tables \ref{tab:stack_measurements}, \ref{tab:stack_fluxes}). However, when converting \ha\ fluxes to SFR applying the \citet{kennicutt_1998} conversion and the reddening correction, the values for cluster and field are formally compatible. In Figure \ref{fig:ms} we show the field and cluster sources in the final stacked samples in the stellar mass-$\mathrm{SFR_{H\alpha},\,SFR_{SED}}$ plane. All the SFRs have been rescaled by a factor $[(1+z)/(1+1.99)]^{2.8}$ to match the cluster redshift. We adopted as reference the MS parametrization given in \citet{sargent_2014}. In the right panel, individual cluster sources seem tentatively more star-forming than the field counterparts, populating the upper envelope of the MS, and the lowest \nha\ ratio corresponds to the highest $\mathrm{SFR_{H\alpha}}$. However, the average properties of the cluster and field populations in the same mass regime are formally compatible, as shown by the stacked values, and this is likely due to the uncertainties on individual $f$ factor estimates. The stacked values are compatible also when considering $\mathrm{SFR_{SED}}$, as shown in left panel of Figure \ref{fig:ms}. However, this may be partly due to the longer timescales probed by the UV stellar emission as a SFR indicator with respect to \ha\ ($\mathrm{t_{UV}}\sim100\,\mathrm{Myr}$, $\mathrm{t_{H\alpha}}\sim10\,\mathrm{Myr}$), which makes $\mathrm{SFR_{SED}}$ insensitive to potentially recent episodes of star formation in cluster sources with respect to the field. For reference, we show in Figure \ref{fig:sfr} the comparison between SED-  and \ha--based SFRs, where $\mathrm{SFR_{SED}}$ for the stacked samples is the mean of single sources values.\\ %The two estimates are in good agreement, not showing any significant systematics: the median $\mathrm{log(SFR_{H\alpha} / SFR_{SED}) }$ value is equal to $0.064$ ($\sim15$\%, well within the $\sim0.2\,\mathrm{dex}$ uncertainty on $\mathrm{SFR_{SED}}$) with a $0.3\,\mathrm{dex}$ scatter. 
In the most conservative approach, considering the uncertainties on the $f$ factor, we cannot fully disentangle the reddening and sSFR (or SFR) effects in producing the observed EW($\mathrm{H}\alpha$) difference. However, reasonably assuming the $f$ factor as independent of the environment and the average $E(B-V)_{\mathrm{cont}}$ values from SED modelling, we can decouple the two effects, ascribing the enhanced observed $\mathrm{EW(H\alpha)}$ in cluster sources to an enhancement in sSFR. In any case, we emphasize how Figure \ref{fig:ew} shows another significant difference between cluster and field SFGs resulting from this work, in addition to the lower gas-phase metallicity.

\section{The Mass-Metallicity Relation}
\label{sec:mzr}

\begin{figure}
\includegraphics[width=0.50\textwidth]{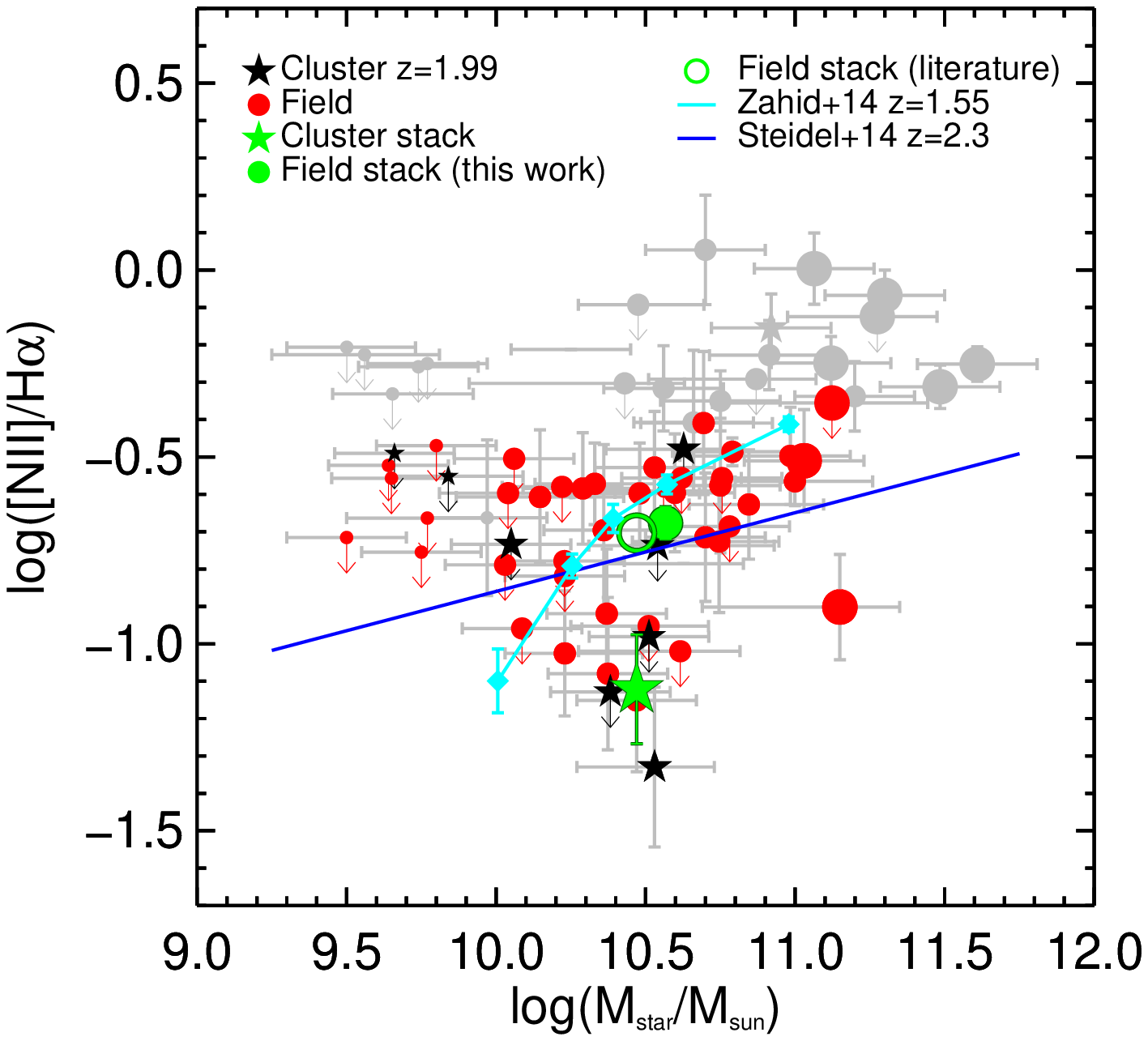}
\caption{Mass-\nha\ relation for the MOIRCS spectroscopic sample. Red circles and black stars represent the field and cluster samples, respectively. Grey symbols mark the objects excluded from the SF sample as AGNs (see Section \ref{sec:line_diagnostics}). Arrows indicate $2\sigma$ upper limits. Symbol sizes scale as the stellar mass. The green solid circle and star represent the field and cluster stacked samples, respectively. The green empty circle marks the expected field position at $z=2$ from the interpolation of literature data (see text for details). Cyan diamonds represent the stacked points from the FMOS survey at $z\sim1.55$ \citep{zahid_2013}. The blue solid line is the relation for the $z\sim2.3$ sample from \citet{steidel_2014} (Equation 17).}
\label{fig:mzr}
\end{figure}
\begin{figure}
  \includegraphics[width=0.50\textwidth]{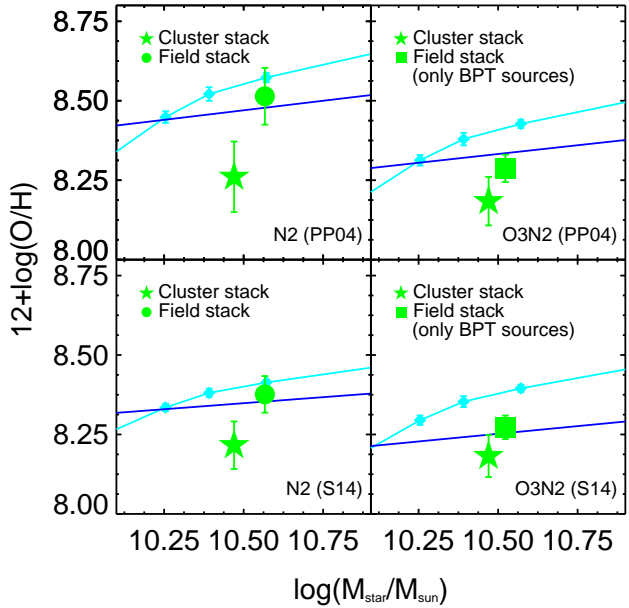}
  \caption{MZR for the MOIRCS spectroscopic stacked samples in the mass range $10 \leq \mathrm{log(M/\msun)} \leq 11$. The green circle and square indicate the $31$--source and $16$--source field stacked values, respectively. The green star indicates the cluster stacked value. Cyan diamonds represent the stacked points from the FMOS survey at $z\sim1.55$ \citep{zahid_2013} and the blue solid line is the relation for the $z\sim2.3$ sample from \citet{steidel_2014} (Equation 17), both rescaled to match the metallicity calibration in each panel (see the legend).}
  \label{fig:mzr_calib}
\end{figure}
The presence of a correlation between stellar mass and metallicity in SFGs has been known for a long time \citep{lequeux_1979}, both locally \citep{tremonti_2004} and at increasing redshift \citep[][and many others]{erb_2006, kewley_2008, zahid_2012, cullen_2014, zahid_2014, steidel_2014, wuyts_2014}. This relation can be interpreted as the result of the interplay among the accretion of metal poor pristine gas, star formation episodes and enriched gas expulsion through stellar winds \citep{dave_2012, lilly_2013, zahid_2014}.
% The final balance of these physical phenomena draws the observed local MZR where more massive galaxies are also more metal rich. 
At higher redshifts the overall observed metallicity is lower than in local galaxies, virtually shifting the observed MZR vertically with redshift. In Figure \ref{fig:mzr} we show the observed \nha\ ratio, a proxy for gas-phase metallicity, as a function of stellar mass. A $>4\sigma$ significant lower ratio is observed in the cluster stack with respect to the field mass-matched sample. This result is unchanged if we consider as a field \nha\ representative value the linear interpolation at $z=1.99$ of the $z\sim1.55$ and $z\sim2.3$ values from \citet{zahid_2013} and S14, at fixed mass. Quantitatively, the metallicity difference between the cluster and field samples depends on the adopted calibration for \nha\, as shown in the left panels of Figure \ref{fig:mzr_calib}. In the same figure we show the metallicity derived from the $O3N2$ indicator (right panels) for the subsample of $16$ intermediate mass field SFGs with \hb\ and \oiii\ measurements. Also in this case, given the comparable \ohb\ values of the cluster and the field samples, the difference in the final metallicity values reflects the different \nha\ ratio through the slope of the adopted linear $O3N2$ calibration -- i.e., through the sensitivity to metallicity variations assigned to \nha. Metallicity differences vary between $0.09\,\mathrm{dex}$ and $0.25\,\mathrm{dex}$ from $O3N2_{\mathrm{S14}}$ and $N2_{\mathrm{PP04}}$ calibrations, respectively. Given the low number statistics, we do not attempt here any fit to the observed points in Figure \ref{fig:mzr}, neither stacked nor single.\\
\begin{figure}
\includegraphics[width=0.50\textwidth]{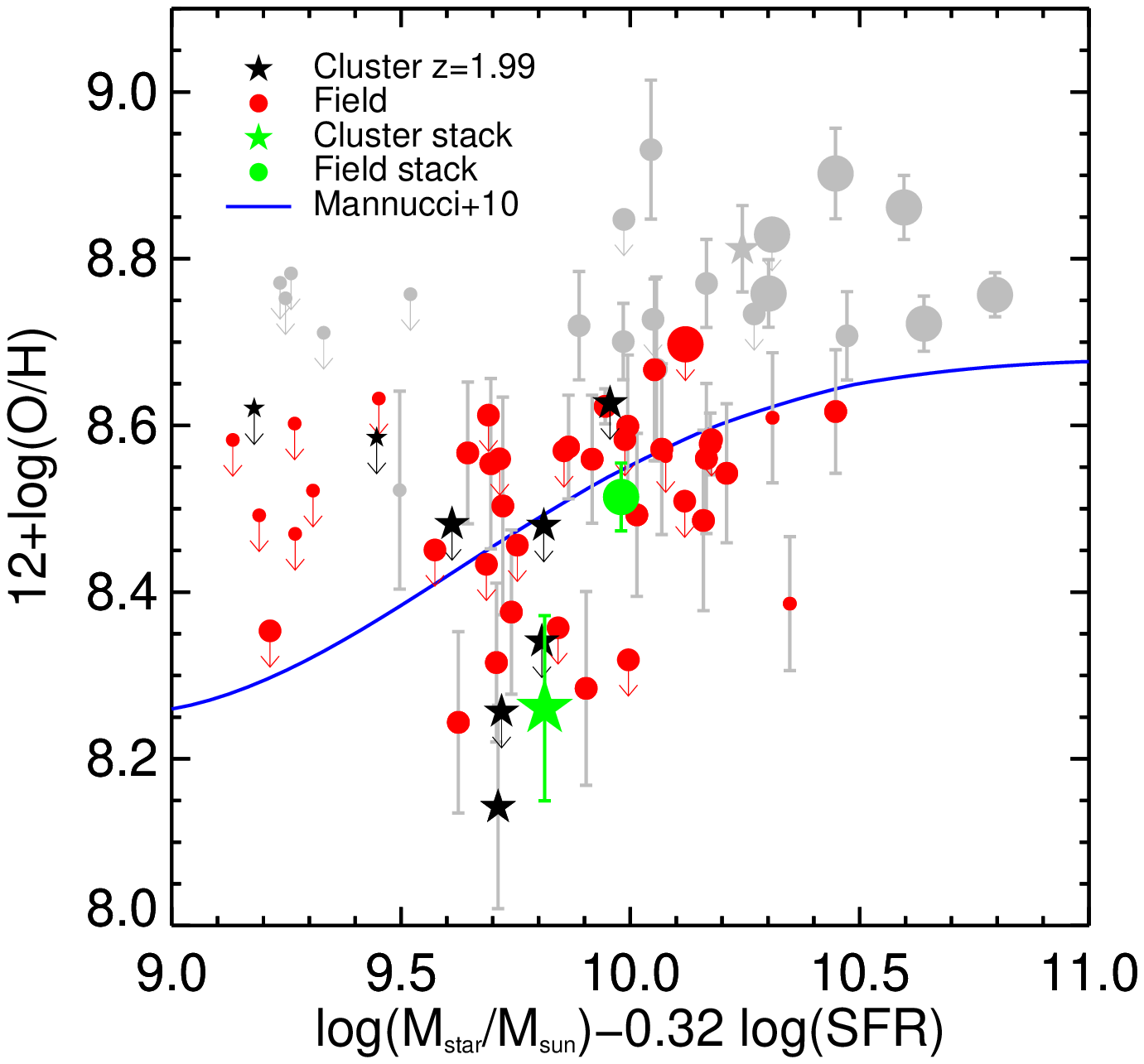}
\caption{FMR for the MOIRCS spectroscopic sample. Red circles and black stars represent the field and cluster samples, respectively.  Grey symbols mark the objects excluded from the SF sample as AGNs (see Section \ref{sec:line_diagnostics}). Arrows indicate $2\sigma$ upper limits. Symbol sizes scale as the stellar mass. The green solid circle and star represent the field and cluster stacked samples, respectively. SFRs are estimated from \ha\ fluxes (see Section \ref{sec:ew_ha} for details). The gas-phase metallicity is estimated from the $N2$ indicator as calibrated by \citet{pettini_2004}. The blue solid line represents the polynomial parametrization of the FMR by \citet[][Equation (4)]{mannucci_2010}.}
\label{fig:fmr}
\end{figure}
Recently, the possible introduction of a third term in the MZR has been advocated to reduce the intrinsic scatter of the relation. \citet{mannucci_2010} proposed to add the SFR to build the so called ``Fundamental Mass-Metallicity Relation'' (FMR) and provided a suitable description of it through the $\mu_{\alpha} = \mathrm{log(M/\msun) - \alpha log(SFR/\msun yr^{-1})}$ parameter. They found that the minimum scatter for their local sample from the SDSS is reached for $\alpha=0.32$, and that this value does not evolve at least up to $z\sim2.5$. This latter finding is somewhat in contrast with recent works at $z\gtrsim1.5$ \citep[S14]{zahid_2014, zahid_2013, wuyts_2014} and, partially, with the results of the present study (but see \citet{maier_2014} for the impact of the choice of the FMR extrapolation on the evolution with $z$). Figure \ref{fig:fmr} shows the FMR projection on the $\mu_{0.32}$-$12+\mathrm{log(O/H)}$ plane as parametrized in Equation 4 of \citet{mannucci_2010}, where we used $\mathrm{SFR_{H\alpha}}$. Again, the choice of the indicator (and especially of its calibration) is decisive for the absolute value of the metallicity which enters the FMR. In Figure \ref{fig:fmr} we show the PP04 $N2$ calibration, which, in the case of the field sample, is consistent with the FMR trend, after a proper conversion from the \citet{maiolino_2008} calibration system to PP04. In the same figure, the cluster value is tentatively inconsistent ($\sim2.7\sigma$) with an unevolving FMR up to $z\sim2.5$. After proper metallicity rescaling, we observe a similar inconsistency with the prediction of the analytically derived $Z(\mathrm{M,SFR})$ by \citet{lilly_2013} (left panel of their Figure 7).

\section{Discussion}
\label{sec:discussion}

\subsection{Potential selection effects}
\label{sec:selection_effects}
\begin{figure*}
  \includegraphics[width=\textwidth]{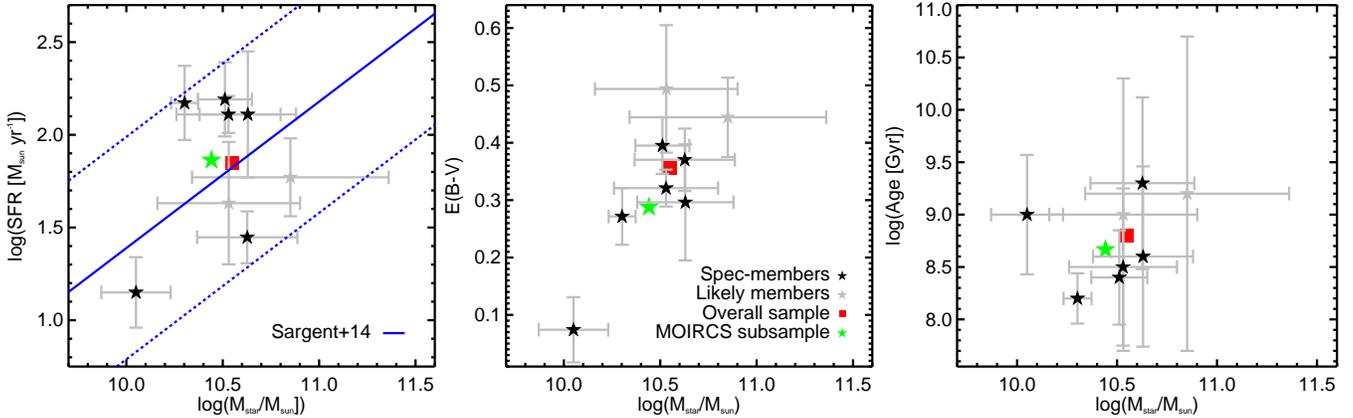}
  \caption{Photometric properties of the cluster ``parent'' sample from which the high priority sample for MOIRCS follow-up has been extracted. In each panel black stars mark the $6$ WFC3 spectroscopically confirmed SF cluster members in the $10 \leq \mathrm{log(M/M_{\odot})} \leq 11$ mass range which have been followed-up with MOIRCS and stacked. Grey stars mark the candidate members in the same mass range not observed with MOIRCS. The green star and red circle indicate the mean value for the sample of followed-up sources and for the overall population, respectively. Left panel: Stellar mass versus SED based SFR. The MS at $z=2$ is represented with a $\pm0.6\,\mathrm{dex}$ scatter as parametrized by \citet{sargent_2014}. Central panel: Mass-Reddening Relation. Right Panel: Mass-luminosity weighted age relation for constant SFH.}
  \label{fig:parent_sample}
\end{figure*}
We checked for possible biases in our field sample comparing it to trends from other surveys at similar redshifts and extrapolating them to $z=1.99$ (Section \ref{sec:mzr}). As shown in Figure \ref{fig:mzr}, our selection of field sources gives results that are consistent with much broader samples in literature \citep[][S14]{zahid_2013}. This shows that our selection is not biased towards specific high redshift galaxy populations, but extracts a representative sample of MS-SFGs at $z\sim2$. This result is confirmed stacking only \textit{sBZK}-, \ha-selected galaxies from the COSMOS mask and comparing them to the general field sample, as we recover fully consistent line ratios (within $1\sigma$ uncertainties).\\
Moreover, for the higher priority assigned to the WFC3-confirmed cluster members over candidates, another possible selection bias could have occurred in the cluster sample. In particular, even if not specifically \oiii--selected, $5/6$ SF cluster members in the final stack have an \oiii\ detection from WFC3, which could have introduced a bias towards the metal poorer cluster members. To check this possibility we investigated the properties of the whole ``parent'' pool of spectroscopically confirmed and candidate star-forming members in the mass range $10 \leq \mathrm{log(M/\msun) \leq 11}$ from which we chose the high priority sample to observe. The mass cut, the constraints on the quality of photometric data, and the SF classification reduced the original pool to $8$ members in the investigated mass bin. $6/8$ are the WFC3 spectroscopically confirmed members that we observed with MOIRCS and which were stacked. The other $2$ sources are candidate members which were not inserted in the final MOIRCS mask because of geometrical constraints in slit positioning. The photometric properties of these $8$ galaxies are shown in Figure \ref{fig:parent_sample}. %From these plots it is clear that the mean properties of the WFC3 spectroscopically confirmed and MOIRCS followed-up population are similar to the ones of the overall population including the two candidate' members, excluding potential biases in the cluster member selection. 
Furthermore we investigated the reasons of the WFC3 non-detection of the $2$ candidate members, checking if the absence of \oiii\ detection could have been due to high metallicities, which could have potentially influenced our subsequent analysis of the cluster metal content. From the SED based SFR and $E(B-V)$ estimates and assuming an intrinsic $\mathrm{H\alpha/H\beta}$ ratio equal to $2.86$, we derived the expected \hb\ observed flux. In both cases it fell well below the WFC3 $3\sigma$ detection threshold ($2\times10^{-17}\,\mathrm{erg\,cm^{-2}\,s^{-1}}$, G13), showing that these $2$ sources are intrinsically faint rather than metal rich (if the latter was the case, we should have detected them in \hb\, but not in \oiii). Moreover, assuming the WFC3 detection threshold, the predicted \hb\ flux, and an empirical track describing the observed population at $z=2$ in the BPT diagram, we estimated the \nha\ ratio for these two sources and the metallicity with $N2$, confirming their homogeneity with the sample of $6$ galaxies that we stacked. This result does not change using tracks describing only cluster sources, the whole sample of $z\sim2$ galaxies, or a trend from literature. We thus conclude that the intrinsic faintness combined to potential high orders contamination in the slitless spectroscopy did not allow a detection and a redshift estimate with WFC3. We show in Appendix the WFC3 G141 spectra for these two sources. In addition we could exclude significant biases introduced by low number statistics for the cluster sample (Section \ref{sec:stacking}). Therefore we can be reasonably confident that evident selection effects are not invalidating the analysis presented in this work. 
% Maybe this can be avoided...
%(MAYBE WE CAN AVOID THIS PART, as I have already commented on Ha normalization in the stacking section. We could remove the plot, too.)
%Finally, although the only cluster source in the final stack with a \nii\ $>2\sigma$-detection has also the highest observed \ha\ flux, we note that the low \nha\ ratio for cluster SFGs seems not to be driven by the brightest source in \ha, since the most stringent \nha\ upper limits are consistent over the observed \ha\ range for the cluster sources in the stack (Figure \ref{fig:halpha_bias}). Moreover, not having implemented a weighting scheme based on the signal-to-noise ratio for the cluster stack, we do not favour the brightest \ha\, highest signal-to-noise ratio spectra over the others.

\subsection{The environmental effect}
\subsubsection{Comparison with other works }
The debate about the environmental signatures in the chemical enrichment of cluster galaxies is still ongoing, even in the local Universe. %In the local Universe, the efforts have been focused on the study of both single galaxies in massive clusters like Virgo and Perseus \citep{skillman_1996, robertson_2012, hughes_2013} and on the ``average'' cluster population \citep{ellison_2009}. \citet{skillman_1996} report a higher metallicity ($\sim0.3\,\mathrm{dex}$) in a limited sample of $9$ gas-deficient SFGs in Virgo, partially confirmed for a sample of $6$ SFGs in the lower density and lower velocity dispersion Perseus cluster \citep{robertson_2012}. The explanation for this result is linked to the effect of ram pressure stripping on SFGs infalling into the clusters, which deprives the galaxies of gas to fuel the SF and allows a metal enrichment from the last generation of formed stars. However, such a substantial deviation in the MZR seems true only for gas-deficient galaxies, for which a clear metallicity-gas-deficiency relation is in place. On the contrary, for the overall population of SFGs in the cluster, the difference is limited to $\sim0.05\,\mathrm{dex}$ \citep{cooper_2008, ellison_2009}, if at all \citep{hughes_2013, mouhcine_2007}. Moreover, such a small offset becomes virtually invisible for unbinned data even in a large sample \citep{ellison_2009}. In absence of a solid gas mass estimate for our sources, we cannot investigate its relation to the oxygen abundance (recently advocated by \citealp{zahid_2014} to find a ``universal'' MZR) and the impact of the environment on this relation. Even the influence of local density - rather than cluster membership - is source of controversy. \citet{mouhcine_2007} do not find a clear dependence on local density over a wide range of values. On the contrary, \citet{ellison_2009} find a higher metallicity ($0.05\,\mathrm{dex}$) for galaxies residing in a local denser environment, almost irrespectively of the cluster membership. 
The situation for high redshift clusters is almost unexplored up to date. \citet{kulas_2013} studied the MZR for a sample of $23$ SFGs belonging to a $z=2.3$ protocluster \citep{steidel_2005}. They find a $0.15\,\mathrm{dex}$ metallicity enhancement for galaxies inside the overdensity with respect to field counterparts at low masses ($\mathrm{log(M/\msun)} \lesssim 10.1$, Chabrier IMF), but no difference at higher masses. Similarly, \citet{shimakawa_2014} find higher gas-phase metallicities in protocluster members than in the field at $z=2.1-2.5$ below $10^{11}\,\msun$. These results are in contrast with the main finding of this work. Unfortunately, we cannot study possible mass trends, given the low number of SFGs in CL J1449+0856 and the high mass limit for completeness. Moreover the mass range that we explored is someway in between the mass bins defined in \citet{kulas_2013}, increasing the difficulty of a direct comparison. Furthermore, we note that contamination from AGNs is potentially an issue in selecting SFGs at high redshift: the inclusion of type-2 AGNs, which could be hosted in a non-negligible fraction of high mass galaxies at $z>1$ \citep[e.g.,][]{trump_2013}, can bias the \nha\ ratio towards higher values and hence their gas-phase metallicities derived from $N2$. \citet{kulas_2013} selected sources according to UV emission, which should prevent strong AGN contamination \citep[see][]{steidel_2014}. \citet{shimakawa_2014} rejected AGNs using a slight modification of the BPT diagram relying on \ha\ fluxes and reddening correction to estimate \hb\ fluxes. In this work we coupled the BPT and \nha--EW($\mathrm{H}\alpha$) diagrams, including X-ray and radio criterion and these different AGN exclusion criteria could have impacted the final results. Furthermore, despite being at comparable redshifts, these overdensities and CL J1449+0856 are structurally different. A pondered definition of \textit{protoclusters} and \textit{clusters} is beyond the scope of this work, but we remark that potentially these different structures may give rise to different effects on their host galaxies, and thus straight comparisons should be made with caution. 
% the higher spatial density, the presence of core red, passive, and massive galaxies, and the $\sim3.5\sigma$ X-ray detection characterising CL J1449+0856 (G11,G13,S13), but not the overdensity examined by Kulas et al. These features make CL J1449+0856 closer to local clusters, even if in its recent past a phase of major assembly occurred, as proven by the massive blue SF population analysed in the present work. Thus we caution against a straight comparison between the two overdensities.    
\subsubsection{The past history of CL J1449+0856: a recent transitional phase of stellar mass assembly?}
\begin{figure}
  \includegraphics[width=0.35\textwidth, angle=90]{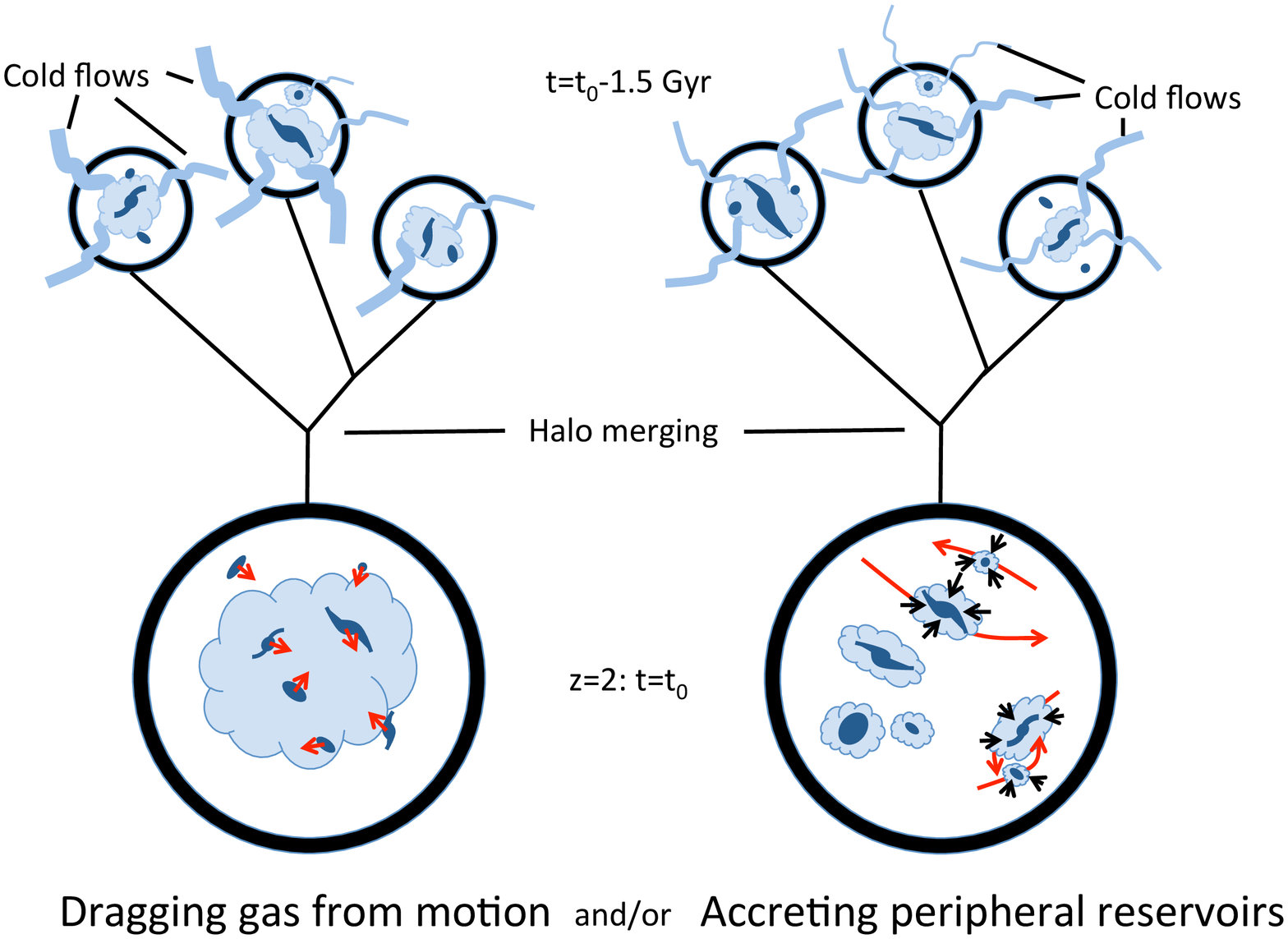}
  \caption{Sketch of the speculative model of gas accretion for SFGs residing in CL J1449+0856. Left branch refers to the possible creation of a gas rich environment in clusters close to a phase of major assembly. Right branch shows the impact of galaxy encounters on the gas halos around each galaxy. The vertical direction marks the time: at the top a phase of gas enrichment occurs $\sim1.5\,\mathrm{Gyr}$ before a major phase of assembly of CL J1449+0856 at $z\geq2$.}
  \label{fig:sketch}
\end{figure}
\begin{figure}
  \includegraphics[width=0.5\textwidth]{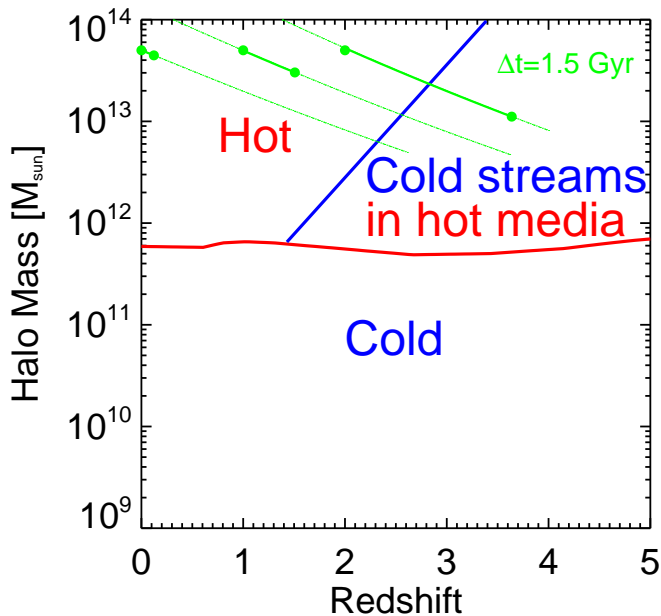}
  \caption{Analytic prediction of the ``hot'', ``cold'', and ``cold streams in hot media'' regimes in the mass-redshift space from \citet{dekel_2009l}. The green tracks show the mass growth for halos of $5\times10^{13}\,\msun$ at $z=0,1,2$ \citep{fakhouri_2010}. The solid segments represent a lookback time interval of $1.5\,\mathrm{Gyr}$ starting from the redshift of reference.}
  \label{fig:dekel}
\end{figure}
\begin{figure*}
  \centering
  \includegraphics[width=0.16\textwidth]{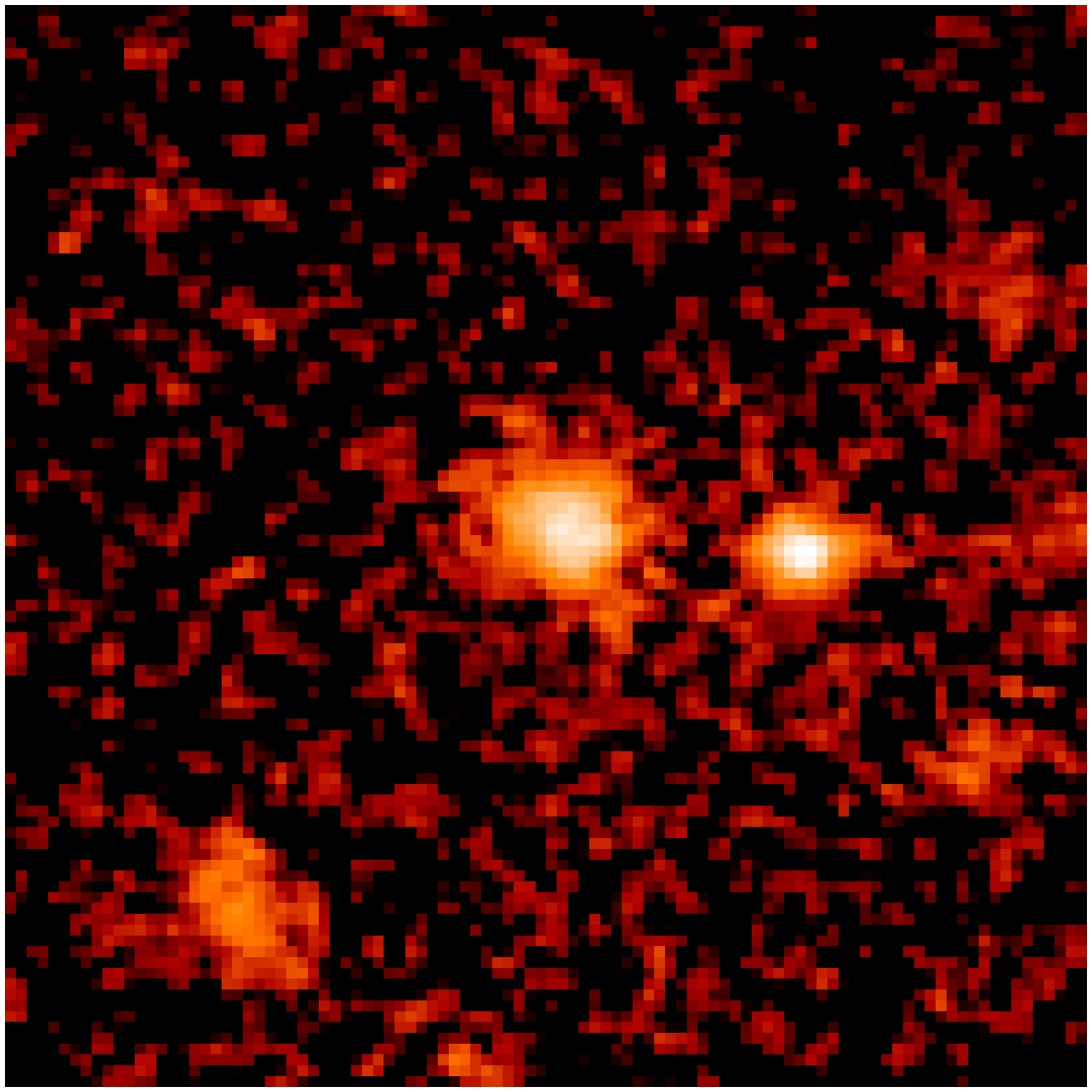}
  \includegraphics[width=0.16\textwidth]{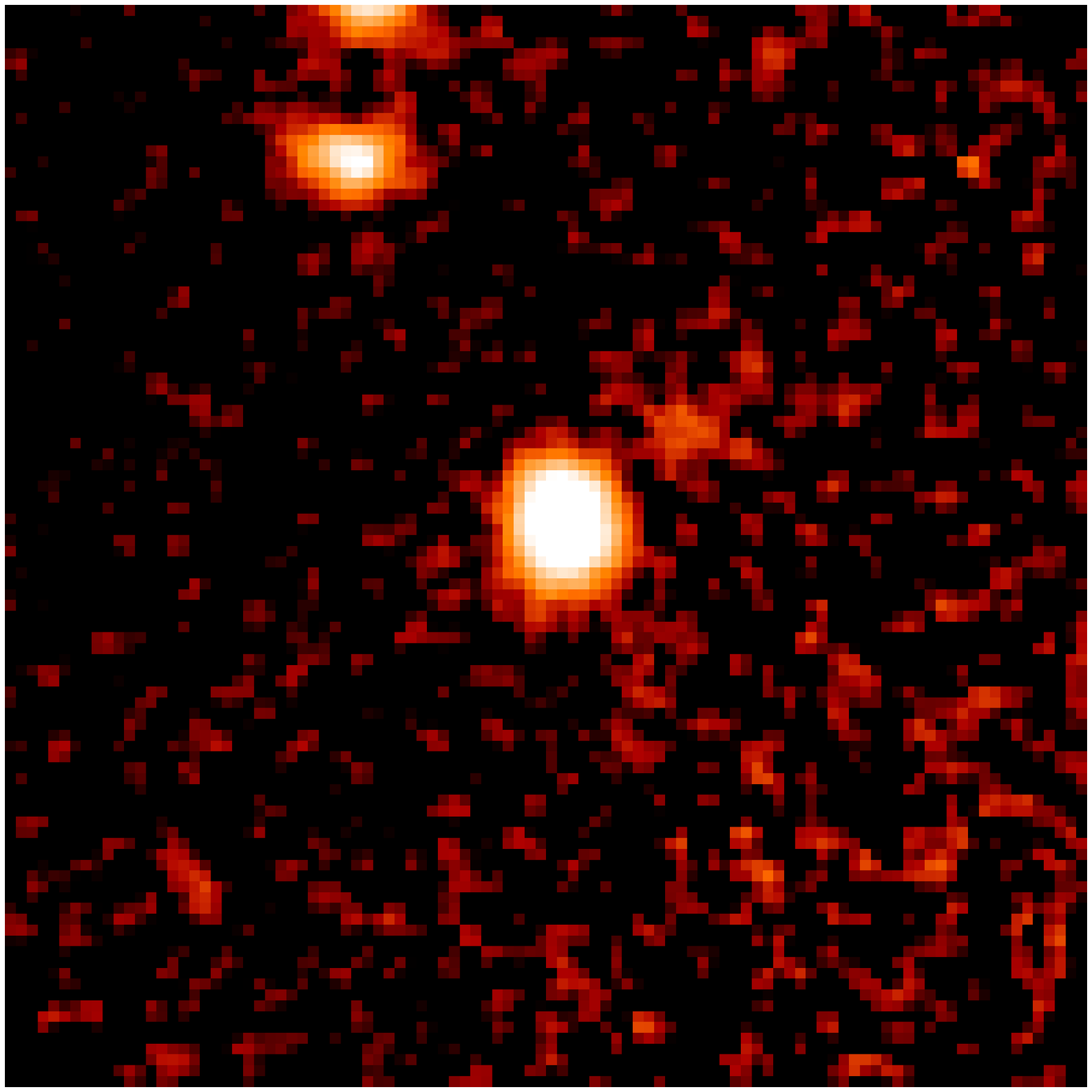}
  \includegraphics[width=0.16\textwidth]{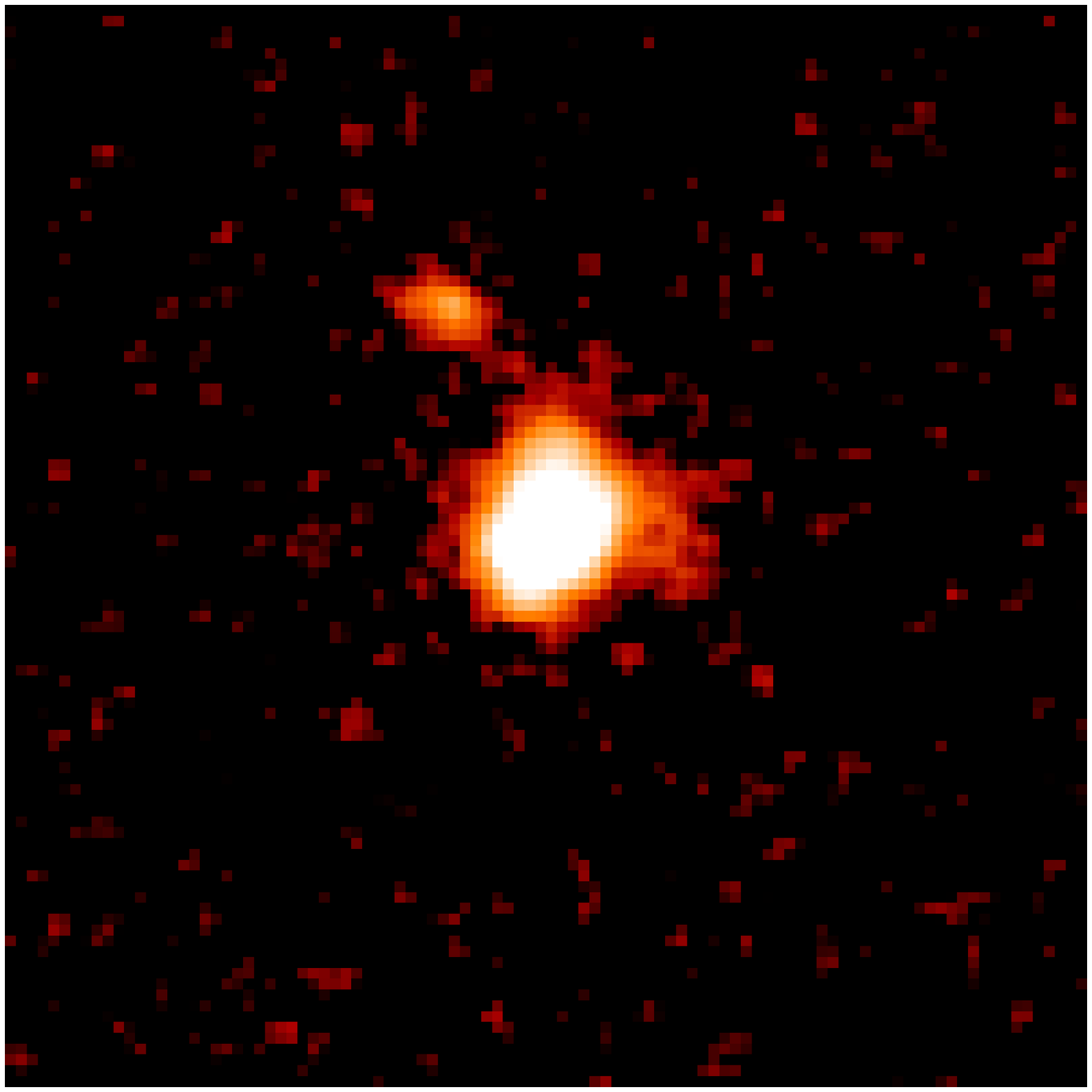}
  \includegraphics[width=0.16\textwidth]{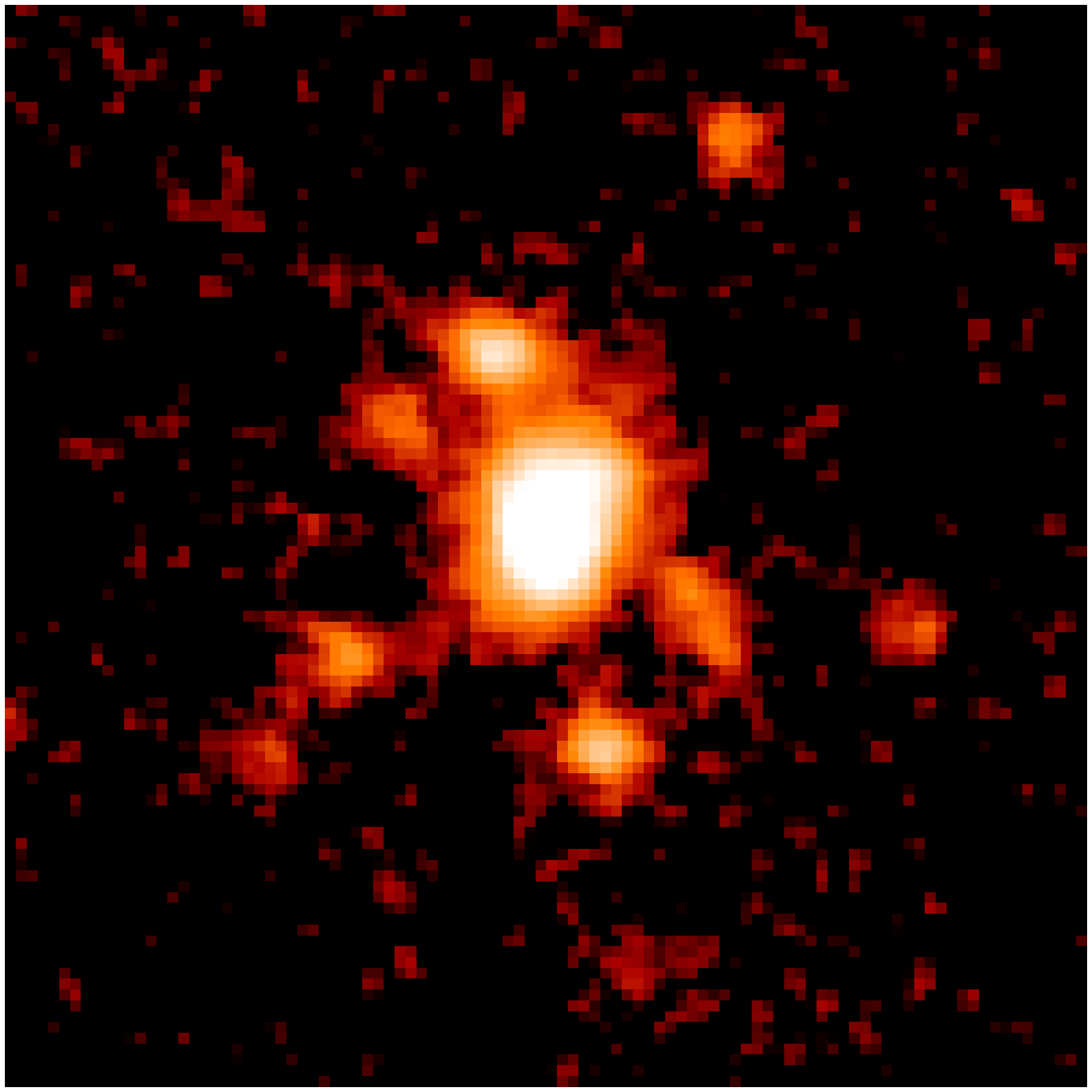}
  \includegraphics[width=0.16\textwidth]{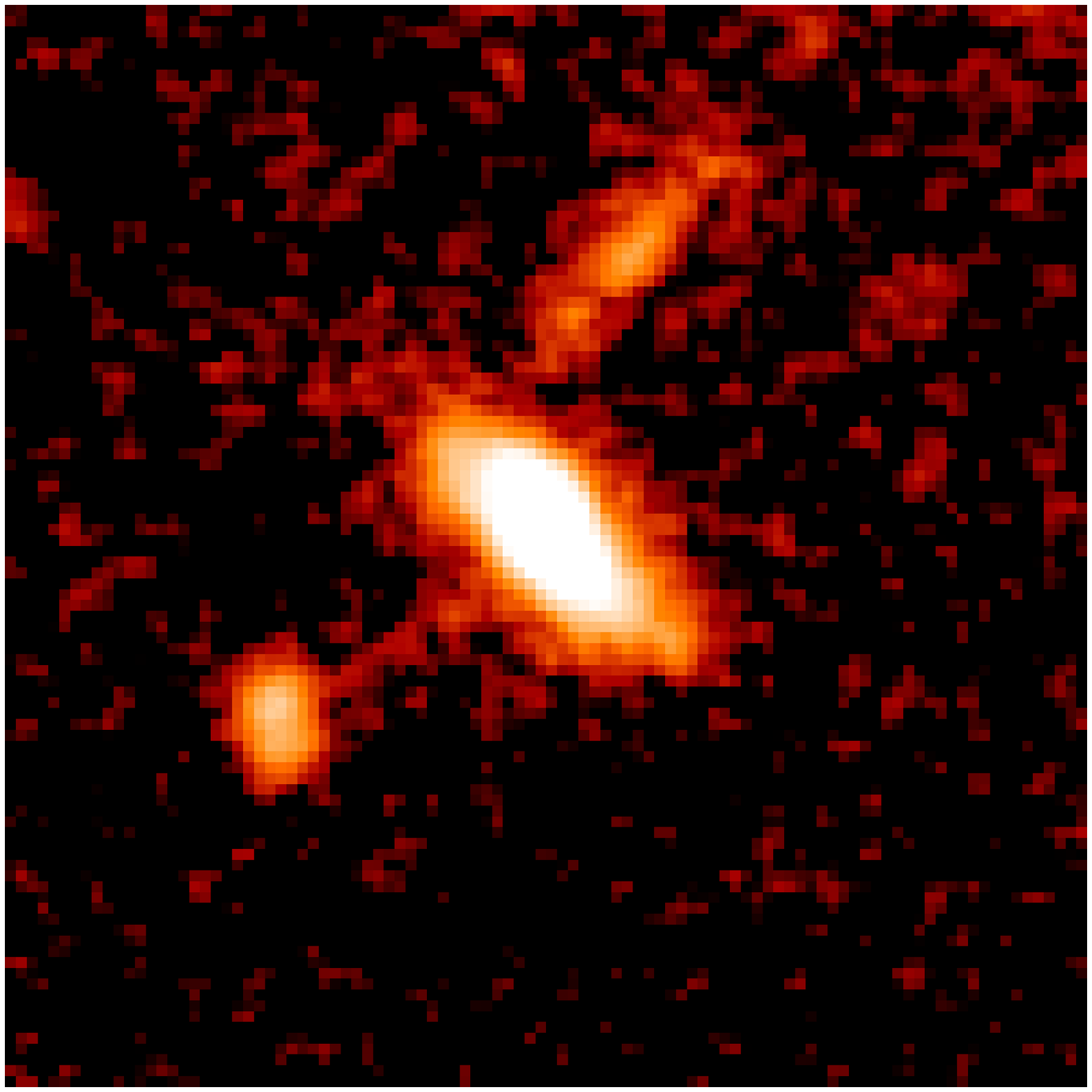}
  \includegraphics[width=0.16\textwidth]{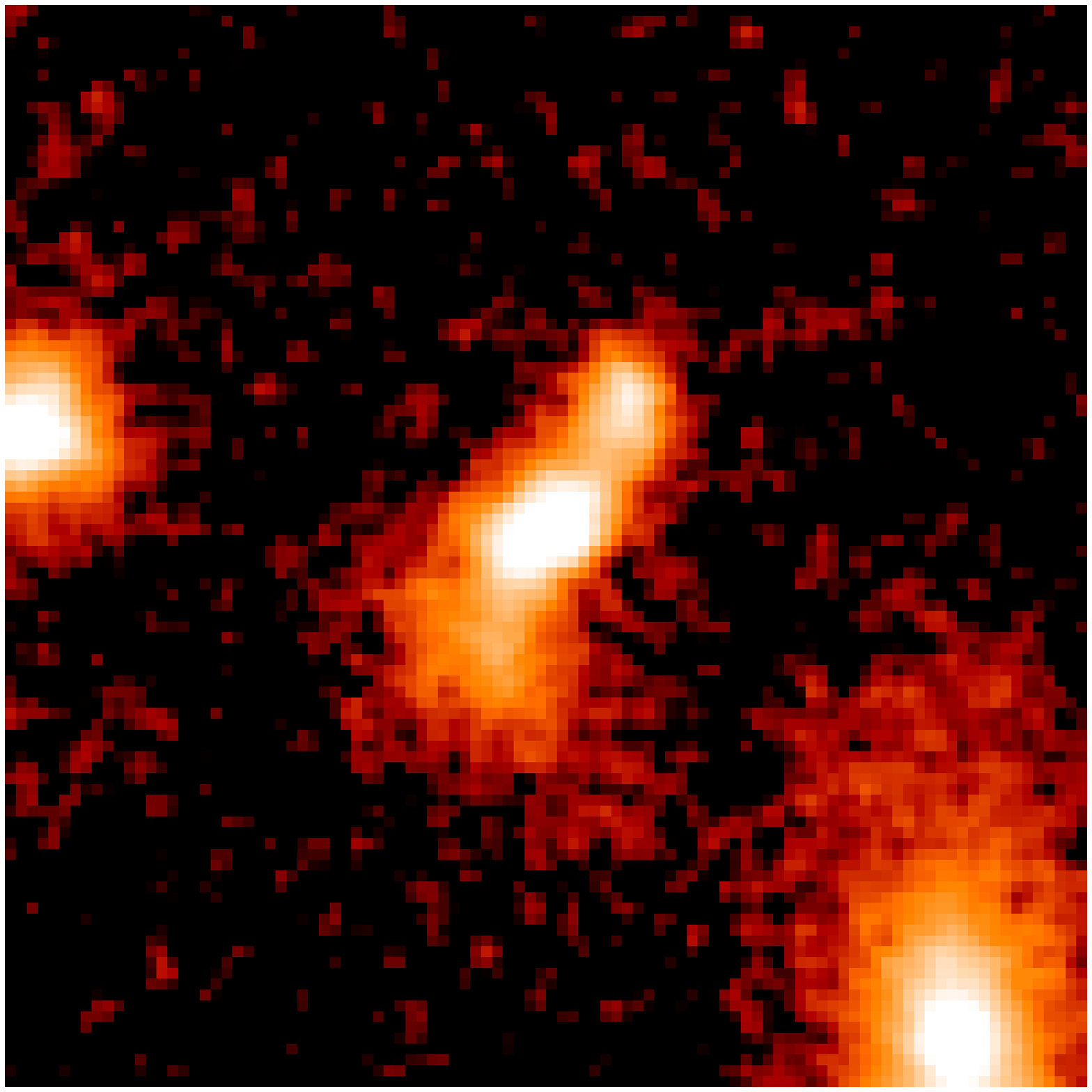}
  \caption{F140W $3"\times3"$ ($\sim25\times25\,\mathrm{kpc}$) cutouts of our sample of cluster SFGs in the mass range $10 \leq \mathrm{log(M/\msun) \leq 11}$ (north is up, east is left).}
  \label{fig:cutouts}
\end{figure*}
The observations presented in this paper have highlighted the presence of a $>4\sigma$ significant difference in \nha\ ratio between a sample of SFGs belonging to CL J1449+0856 and a mass-matched sample in the field. This difference is directly translated in a metallicity difference with all the indicators employed in the analysis, so that cluster sources are a factor $0.09$-$0.25\,\mathrm{dex}$ (using $O3N2_{\mathrm{S14}}$ and $N2_{\mathrm{PP04}}$, respectively) more metal poor than the field counterparts. What follows is a speculation about the origin of this effect. As discussed above, CL J1449+0856 is partially virialized, but a relatively recent phase of assembly must have occurred. According to the model of halo mass growth by \citet{fakhouri_2010}, a halo of $5\times10^{13}\,\msun$, such as the one hosting CL J1449+0856 (from X-ray emission, G13), should have increased its mass of a factor $\times2$ ($\times5$) in the previous $\sim1\,\mathrm{Gyr}$ ($\sim1.5\,\mathrm{Gyr}$) of its lifetime. The recent coalescence of multiple less massive halos could have impacted the hosted galaxies in a twofold way, graphically sketched in Figure \ref{fig:sketch}. First, the single bricks forming the final halo could have been gas enriched through cold streams and subsequently merged, thus creating an environment rich in pristine gas. The accretion of cold gas even in quite massive halos at redshifts close to the formation epoch of CL J1449+0856 progenitors is consistent with model predictions, if high density, steady, cold streams penetrating the shock heated medium are considered \citep{birnboim_2003, keres_2005, brooks_2009, dekel_2009p, dekel_2009l}. In Figure \ref{fig:dekel}, the mass growth track for CL J1449+0856 halo \citep{fakhouri_2010} crosses the line separating the hot ISM and cold streams in hot ISM regimes $\sim 1\,\mathrm{Gyr}$ before $z=2$ and thus the progenitors of the cluster halo could have been recently enriched of cold gas, before merging. If this gas is not prevented from cooling \citep{fabian_1994, revaz_2008, salome_2011}, given the high dark matter density in $z\gtrsim2$ halos, it might be dragged and accreted on the galaxies simply moving across it, diluting the metal content. This first effect might not be effective at lower redshifts, where the gas reservoirs in halos are at lower densities, prevented from cooling after the cluster full virialization, chemically enriched by Gyrs of stellar formation, and not replenished by cosmological inflows. This latter aspect is illustrated in Figure \ref{fig:dekel}, where the mass growth tracks for a halo of $5\times10^{13}\,\msun$ at $z=0,1$ do not enter the region of cold streams in hot media on timescales of the order of $\sim1\,\mathrm{Gyr}$. Moreover, in a $\Lambda\mathrm{CDM}$ universe the baryon growth rate scales as $(1+z)^{2.25}$ at fixed mass \citep{neistein_2008}. Thus, at low redshifts the progenitors of a halo of such mass cannot be easily refurnished of cold gas.\\
Secondly, a recent epoch of high merging rate of dark matter halos could have favoured encounters, fly-bys, and mergers among the galaxies hosted in the merging halos, given the low (but increasing) cluster velocity dispersion. An encounter can trigger the accretion of the reservoirs of cold, rich, and pristine gas located in the halos of single galaxies at $z\gtrsim2$, continuously replenished by cosmological inflows \citep{ceverino_2010, gabor_2014}. The accretion of such gas would lower the metallicity, as observed in our sample, and subsequently enhance the SF. In addition galaxy minor and major mergers could lower the metal abundance themselves \citep{contini_2012, queyrel_2012}. Moreover, the final cluster potential well in which the galaxies reside and interact can facilitate the merging events through the so called ``gravitational focusing'' effect \citep{martig_2007, moreno_2013}, accelerating the gas accretion from the galaxy outskirts. Recent observational studies have shown a SFR increase and metallicity decrease (up to $\sim0.07$-$0.09\,\mathrm{dex}$) in close pairs and post-mergers in the local Universe \citep{ellison_2013}. This result is supported by simulations suggesting that mergers induce the funneling of gas reservoirs from the peripheric regions of galaxies towards the center, diluting the metallicity and triggering new SF \citep{torrey_2012}. This second effect could be effective in terms of gas accretion on galaxies entering the halo of low redshift clusters, generating a SFR enhancement in the cluster outskirts, as observed for example in Virgo \citep{temporin_2009}. However, given the chemical enrichment due to stellar formation in the last $10\,\mathrm{Gyr}$, the gas accretion might not be effective in reducing the metallicity at low redshift. Deep F140W images of our sample of cluster SFGs in the final stack are shown in Figure \ref{fig:cutouts}. Every object shows a disturbed morphology and/or a close companion, which might be hint of high gas fractions or a close encounter, even if the lack of a redshift determination for the companions does not allow to draw a robust conclusion (see, e.g., \citealt{zanella_2015} for the specific case of ID568, Figure \ref{fig:cutouts}, fifth panel). We defer to a future work the detailed study of galaxy morphologies in CL J1449+0856 and a proper comparison with a morphologically characterized field sample.\\   
The transitional epoch that we have just described could be a key phase for galaxy clusters with the assembly of a substantial fraction of stellar mass in SFGs. In a time interval of $500\,\mathrm{Myr}$, typical doubling time at $z=2$ \citep{daddi_2007} and typical scale of gas consumption in SFGs \citep{daddi_2010}, each galaxy in our cluster sample would form stars for a total of $\sim3-6 \times  10^{10}\,\msun$, given the average SFR we measure ($\mathrm{SFR_{H\alpha}} = 112\,\msun\,\mathrm{yr^{-1}}$, $\mathrm{SFR_{SED}} = 73\,\msun\,\mathrm{yr^{-1}}$). This would double the stellar mass already present in cluster SFGs and increase the overall cluster stellar mass in spectroscopically confirmed members by $\sim15$-$35$\% in this time interval or, if we extend this reasoning to the past history of the cluster, SFGs could have assembled an important part of the total stellar mass in a relatively short period of time during this phase.\\
We can gain physical insight about the metal deficiency for cluster SFGs with a simple computation. To explain a $0.15\,\mathrm{dex}$ metallicity difference (a factor $\sim40$\%), given that $\mathrm{M_{\star} \simeq M_{gas}}$ in MS-SFGs at $z\sim2$ \citep{bouche_2007,daddi_2008}, we would need a mass of accreted pristine gas $2\times10^{10}\,\msun$ to dilute the metal content of each galaxy. If we assume the presence in the cluster halo of a gas mass free to cool down and to be dragged and accreted by SFGs in their motion equal to $\sim15$\% of the total halo mass ($\mathrm{M_{halo}} = 5\times 10^{13}\,\msun$), the mass accretion rate would be $35\,\mathrm{\msun\,yr^{-1}}$, assuming a gravitational focusing term $\simeq 5$, a velocity dispersion of $500\,\mathrm{km\,s^{-1}}$, $\mathrm{R_{200}} = 0.4\,\mathrm{Mpc}$, and $4\,\mathrm{kpc}$ as a typical SFG radius ($\mathrm{\dot{M}_{acc} =  \rho_{gas,halo} \cdot \pi R_{200}^2 \cdot v_{disp} \cdot f_{grav} ~ [\msun\,yr^{-1}]}$). However, if we consider only the highest density regions at the core of the cluster ($\mathrm{R}_{\mathrm{clu}} \sim 200\,\mathrm{kpc}$), where the gas is likely to collect, the accreted gas mass could rapidly increase by a factor $\sim8$, enough to halve the metal content of the galaxy. The complementary mechanism linked to galaxy encounters could provide an extra gas accretion rate of $\geq 45\,\mathrm{\msun\,yr^{-1}}$, considering the galaxy density within the cluster ($\sim75\,\mathrm{Mpc^{-3}}$ within $\mathrm{R}_{200}$) and a typical distance for a fly-by of $50\,\mathrm{kpc}$ ($\mathrm{\dot{M}_{acc} = M_{res} \cdot n_{coll} \cdot f_{grav} ~ [\msun\,yr^{-1}]}$, where $\mathrm{n_{coll}}$ is the collision rate and $\mathrm{M_{res}}$ the galaxy halo gas reservoir). The reservoirs available in the galaxy outskirts are expected from simulations to be a few $10^{9}\,\msun$ within a $15\,\mathrm{kpc}$ radius around the galaxy \citep{ceverino_2010,gabor_2014} and they can be replenished only as long as cold inflows can reach the galaxy, which may not be true once the galaxy enters deeply in the cluster halo. Moreover, we cannot rule out the possibility that the local density may be a more important driver than cluster membership (i.e., large scale environment). Unfortunately the very low number statistics of cluster SFGs does not allow for a proper comparison among objects within and outside the virial radius to check for the effective influence of the underlying overdensity on the metallicity, neither considering stacked spectra. Future spectroscopic follow-up of the remaining population of spectroscopically confirmed and candidate SF members will be decisive to clarify this complex picture.    

\section{Summary and conclusions}
\label{sec:conclusion}
We have presented the results of the MOIRCS near-IR spectroscopic follow-up of the SF population residing in CL J1449+0856 at $z=2$. Adding the pre-existing 13-bands photometry of the field and the deep grism G141 slitless spectroscopy of WFC3, we studied the properties of our sample of cluster SFGs in the mass range $10 \leq \mathrm{log(M/\msun) \leq 11}$ with respect to a mass-matched field sample at comparable redshifts through stacking. In our analysis we showed that:

\begin{itemize}
\item the field and cluster samples of SFGs in the studied mass range show comparable \ohb\ ratios, but a $\sim4\sigma$ significant difference in \nha\ ratios. Using different calibrations of the $N2$ and $O3N2$ metallicity indicators, the lower \nha\ ratio measured in cluster SFGs is translated in a $\sim0.09-0.25\,\mathrm{dex}$ (using $O3N2_{\mathrm{S14}}$ and $N2_{\mathrm{PP04}}$, respectively) metal deficiency for the objects belonging to the overdensity. The low metallicity value in cluster sources is confirmed using $R_{23}$ and $O_{32}$ indicators. Furthermore it is supported by the low N/O ratio that we measured ($\mathrm{log(N/O)} = -1.18\pm0.15$). The ionization parameter in the cluster stacked sample from $R_{23}$, $O_{32}$ ($\mathcal{U} \simeq -2.61$) is higher than typical values for local galaxies, but consistent with other determinations at high redshift  
\item We observe $\sim4.7\sigma$ significant $2.5\times$ higher \ha\ luminosity and $\mathrm{EW(H\alpha)}$ in the cluster stack, likely due to enhanced sSFR, even if lower dust reddening and/or an uncertain environmental dependence of the continuum-to-nebular emission differential reddening $f$ may play a role. Thus the metal deficiency observed in the cluster sources appears to be correlated with an increase in the SFR with respect to the field; however we report a $\sim2.7\sigma$ inconsistency with the prediction of a FMR not evolving up to $z=2.5$. 
\item the nebular lines reddening at $z\sim2$ is $\sim1.4\times$ higher than that of stellar continuum estimated through SED fitting, lower than the previous estimates from local measurements and in agreement with recent studies at $z\geq1.5$ \citep{kashino_2013,pannella_2014} 
\item our sample of high redshift galaxies are offset from the local SF sequence on the BPT emission-line diagnostic diagram. This result is in agreement with previous studies at similar redshifts \citep[][and others]{erb_2006, yabe_2012, zahid_2013, steidel_2014}, pointing towards a possible evolution with redshift of the physical conditions of the line emitting regions
\item The metal deficiency in this $z=1.99$ cluster could be due to the accretion of pristine gas which might have diluted the metal content. We speculate that the accretion of large galactic scale gas reservoirs facilitated by the gravitational focusing effect may be responsible for the observed low metal abundance in star-forming cluster members.
\end{itemize}     
%Up to date the lack of proper spectroscopic characterization of galaxies in clusters at $z>1.5$ and their intrinsic low number do not allow to draw strong conclusions. In addition, the poor knowledge of the physical conditions of the ionized ISM at these redshifts makes the effort of distinguishing true environmental effects from internal regulated processes even harder. A proper extensive spectroscopic follow-up with new generation near-IR spectrographs will clarify this picture in the future.      

\acknowledgements
We acknowledge the constructive comments from the anonymous referee, which significantly improved the content and presentation of the results. FV, ED, and AZ were supported by grants ERC-StG UPGAL 240039 and ANR-08-JCJC-0008. FB and SJ acknowledge support from the European Research Council through grant ERC-StG-257720. NA has been supported by the Grant-in-Aid for the Scientific Research Fund under Grant No.23224005. This work is based on data collected at Subaru Telescope, which is operated by the National Astronomical Observatory of Japan.
%\clearpage
\bibliography{paper_bibliography}

\clearpage
\appendix
\section{Appendix A}
As noted in Section \ref{sec:stacking}, averaging single spectra does not necessarily coincide with averaging spectral derived quantities. The difference between these two averaged trends depends on the relationship between the the fluxes of single lines and the derived quantities. In our case we have evaluated the impact of this difference on the mean metallicity calculated through the strong line ratio $\mathrm{[NII]/H\alpha}$ for a population of MS SFGs. Assuming a functional form for the $\mathrm{M_{\star}}$-SFR relation and the $\mathrm{H\alpha}$-SFR conversion, one can easily convert the stellar mass of a galaxy into its intrinsic \ha\ luminosity and, given a Mass-Reddening Relation, into the observable \ha\ flux at a fixed redshift. For this exercise we have used \citet{sargent_2014} MS parametrization as a function of redshift and the standard \citet{kennicutt_1998} relation to pass from \ha\ intrisic luminosities to SFRs. As a MRR we have used the observed trend of our overall sample of SFGs given by the SED fitting described in Section \ref{sec:SED_fitting} and shown in Figure \ref{fig:mrr}. Then we can convert the stellar mass into the gas-phase metallicity using a parametrization of the MZR (or of the FMR if we want to include the effect of the SFR). Here we have used the \citet{zahid_2014} parametrization, given its simple form. Finally we need a conversion from gas-phase metallicities to observed line fluxes. We adopt here the \citet{pettini_2004} calibration of the \nha\ ratio, but in principle we could test any other strong-line ratio. All these relations are somehow scattered and we have adopted the quoted scatters to introduce a gaussian random noise to make our simple simulation more realistic. Given all these relations, we simply generate a random sample of masses in an interesting mass range and derive two estimates of the average metallicity: first we simply compute the average of the single metallicities in mass bins as obtained from the MZR ($Z_{\mathrm{av}}$); then we compute the mean metallicity in the same mass bins as derivable from an hypothethic stacking of the spectra of single galaxies, namely from the average of single line fluxes ($Z_{\mathrm{stack}}$). Analitically it can be shown that $Z_{\mathrm{av}}$ and $Z_{\mathrm{stack}}$ depend on \nii\ and \ha\ fluxes in different ways, so that a priori they can be different:
\begin{equation}
\mathrm{log} \left[  \frac{Z_{\mathrm{av}}}{Z_{\mathrm{stack}}} \right] = \mathrm{log \left( \frac{1}{N} \right)  }  + \mathrm{log} \left[  \sum_{\mathrm{i}=1}^{\mathrm{N}} \left( \mathrm{\frac{[NII]}{H\alpha}} \right)_{\mathrm{i}}^{0.57} \right]- \mathrm{log} \left[  \left(\frac{\mathrm{\sum_{i=1}^{N}}\mathrm{[NII]}_{\mathrm{i}}}{\mathrm{\sum_{i=1}^{N}}\mathrm{H\alpha}_{\mathrm{i}}} \right)^{0.57} \right]
\end{equation}       
where $\mathrm{N}$ is the total amount of observed galaxies and $0.57$ is the slope of PP04 calibration. Given the parametrizations we adopted, the difference between the two mean estimates decreases with increasing stellar mass. Moreover, averaging the metallicity in smaller mass bins gives rise to smaller differences between $Z_{\mathrm{av}}$ and $Z_{\mathrm{stack}}$. Finally, a high number of observed points is more robust against the scatter of the relations we used, reducing the possibility to find huge $\mathrm{log} (Z_{\mathrm{av}}/Z_{\mathrm{stack}})$ ratios in a mass bin. We have stacked sources in a limited high mass regime ($10 \leq \mathrm{log(M/\msun)} \leq 11$) where we have a relatively low number statistics for the cluster sample ($N \sim 10$) and a fairly more robust sample of field galaxies ($N \sim 30$). For a simulated sample of $10$ galaxies at $z = 2$ in the mass bin $10 \leq \mathrm{log(M/\msun)} \leq 11$ the median difference is $ \mathrm{log} (Z_{\mathrm{av}}/Z_{\mathrm{stack}}) = 0.018$ ($\sim4$\%) with a semi-interquartile range of $0.022\,\mathrm{dex}$ over $1000$ runs of the simulation. At the same redshift and mass bin but for $N = 30$ observed points, the median difference is reduced to $ \mathrm{log} (Z_{\mathrm{av}}/Z_{\mathrm{stack}}) = 0.008$ ($\sim1$\%) with a semi-interquartile range of $0.005\,\mathrm{dex}$. At these masses the impact of adopting one approach or the other is restrained, but it could be more important at lower masses - due to the steeper MZR -, at which the stacking technique is usually widely used.

\section{Appendix B}
\begin{figure}
  \centering
  \includegraphics[width=0.35\textwidth, angle=90]{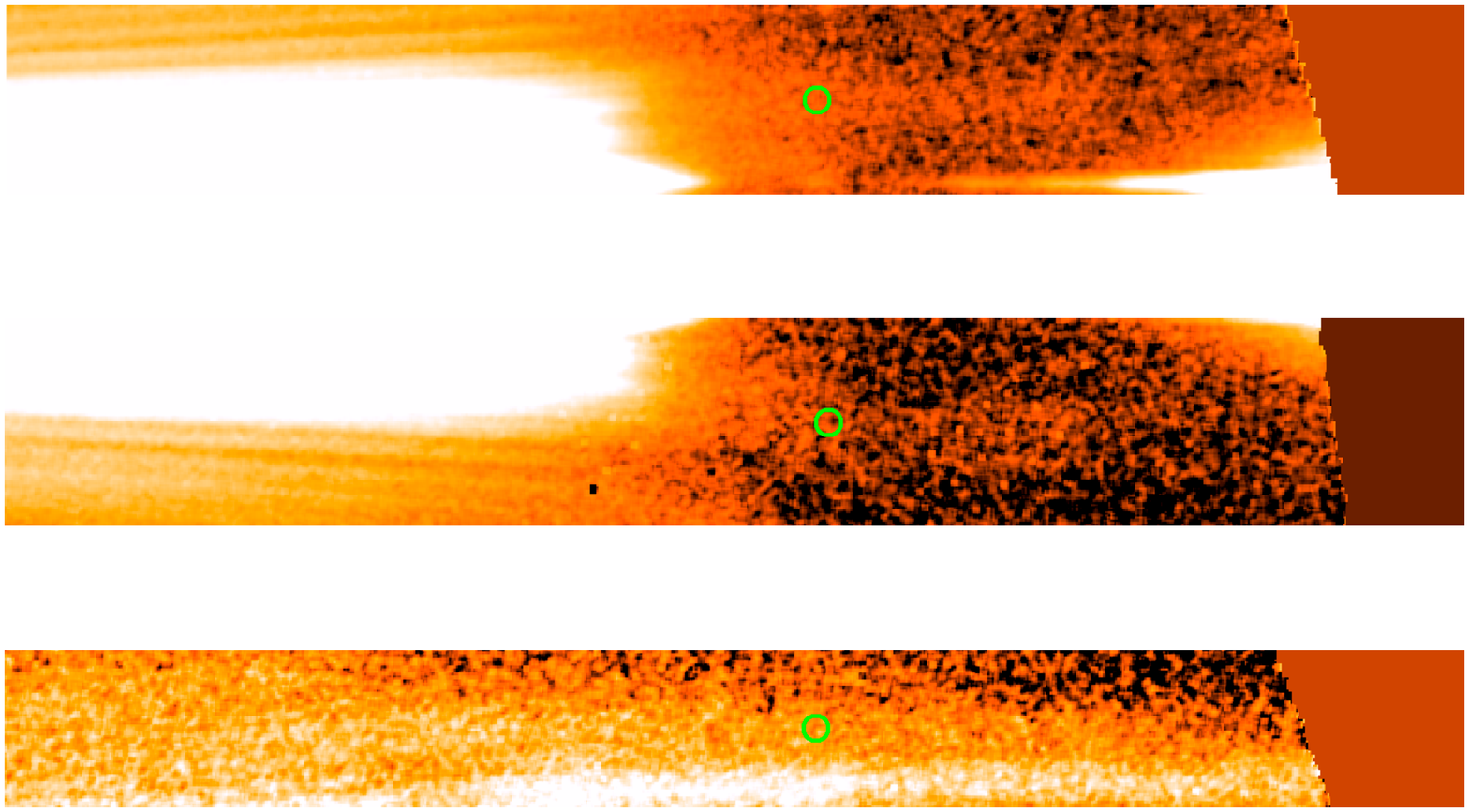}
  \includegraphics[width=0.35\textwidth, angle=90]{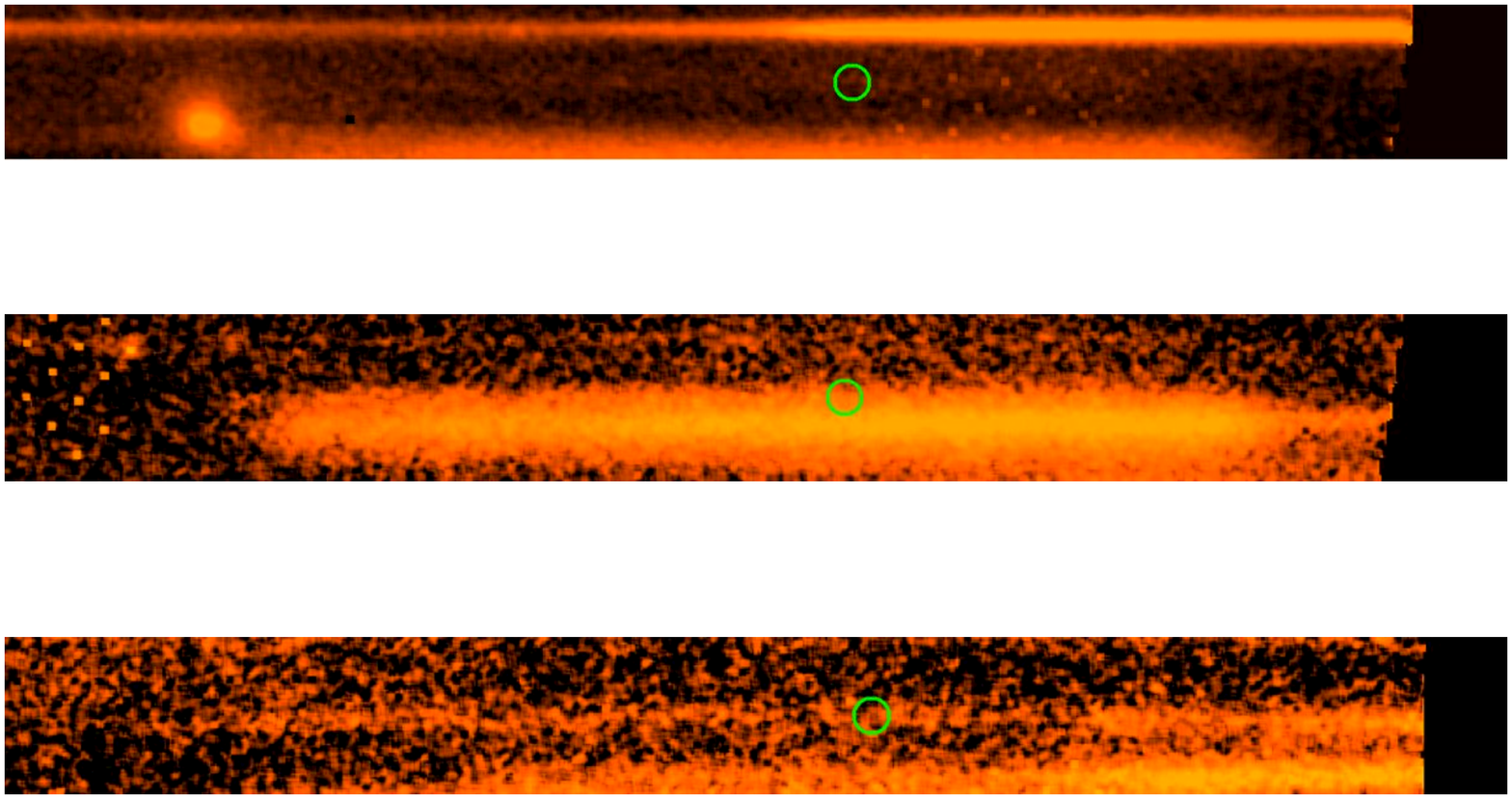}
  \caption{WFC3 G141 slitless spectra of the two cluster candidate members with masses $10 \leq \mathrm{log(M/\msun) \leq 11}$ that were not observed with MOIRCS. Left column: ID 424. Right column: ID 660. The three spectra correspond to separate \textit{HST} visits (G13). Green circles mark the expected position of \oiii\ if the sources were at $z=1.99$}
  \label{fig:wfc3_spec}
\end{figure}

In Figure \ref{fig:wfc3_spec} we show the WFC3 G141 slitless spectra for two candidate members with masses $10 \leq \mathrm{log(M/\msun) \leq 11}$ that were not inserted in MOIRCS masks because of geometrical constraints in slit positioning. Contamination from high orders severily affected these spectra in two of the three \textit{HST} visits, limiting the usable integration time to 4/18 orbits and leading to a higher detection threshold. This did not allow us to detect these intrinsically faint objects (see Section \ref{sec:selection_effects}).

\end{document}